\documentclass[reprint,amsmath,amssymb,aps,nofootinbib,prd]{revtex4-2}

\usepackage{graphicx}
\usepackage{dcolumn}
\usepackage{bm}
\usepackage[dvipsnames]{xcolor}
\usepackage[normalem]{ulem}
\usepackage{natbib}
\usepackage{enumerate}
\usepackage{makecell}

\newcommand{\aap}{Astron.\ Astrophys.}
\newcommand{\mnras}{Mon.\ Not.\ R.\ Astron.\ Soc.}

\newcommand{\apjl}{Astrophys.\ J.\ Lett.}

\newcommand{\jcap}{J. Cosmology \& Astropaticles.}

\newcommand{\Msun}{M_\odot}
\newcommand{\Mmax}{M_\text{\rm\textsc{tov}}}
\newcommand{\Rmax}{R_\text{\rm\textsc{tov}}}
\newcommand{\csmax}{c_\text{\rm s\textsc{tov}}}
\newcommand{\rhomax}{\rho_\text{\rm\textsc{tov}}}
\newcommand{\Pmax}{P_\text{\rm\textsc{tov}}}
\newcommand{\Rhalf}{R_\text{1/2}}
\newcommand{\ratio}{r_\text{1/2}}

\newcommand{\paperi}{Paper~I}
\newcommand{\basic}{\textit{main}}
\newcommand{\mon}{\textit{min+masses}}
\newcommand{\ron}{\textit{min+radii}}
\newcommand{\mimimi}{\textit{minimal}}

\begin{document}
 

\title[Sensitivity Study of NS EoS Inferences]{Sensitivity of the Neutron Star Equation of State Inferences to Mass and Radius Measurements}%

\author{Dmitry D. Ofengeim$^1$}\email{Corresponding Author: ddofengeim@gmail.com}
\author{Peter S. Shternin$^2$}
\author{Tsvi Piran$^1$}%

\affiliation{$^1$Racah Institute of Physics, The Hebrew University, Jerusalem 91904, Israel}
\affiliation{$^2$Ioffe Institute, Politekhnicheskaya 26, St. Petersburg, 194021, Russia}

\date{\today}

\begin{abstract}
We examine how inferences of the neutron-star equation of state  depend on mass and radius observations. We update previous results with recent 
measurements combined with theoretical input from chiral effective field theory and perturbative quantum chromodynamics. The revised constraints are consistent with, but tighter than, those obtained in earlier work.
Isolating the effects of different classes of observations 
we find that the theoretical constraints, together with the requirement that the maximal neutron-star mass exceeds  $\sim 2\,M_\odot$, dominate the equation-of-state inference over most densities. Radius measurements mainly 
refine the constraints at the low-density regime, $\rho \lesssim 2\rho_0$, whereas measurements of masses well above $2\,\Msun$ improve the constraints over a wider density range. 
Finally, we explore the impact of possible future observations. 
The largest impact would  arise  from a measurement that refines the value of the maximal neutron star mass.
It can be, e.g., a detection of an extremely massive neutron star or an improved upper limit. 
However, even a precise measurement a $2.5-2.6\,\Msun$ NS will not alter our knowledge of the equation of state qualitatively.
Conversely, observations lying well outside the present allowed region, would point to new physics in neutron-star cores and require a revision of the current framework.
\end{abstract}

\maketitle

\section{Introduction}
\label{sec:intro}

Determining the neutron star (NS) equation of state (EoS) is a current major challenge in both physics and astrophysics. Significant efforts have been devoted to this problem from both theoretical and observational perspectives. Matter in the deepest NS layers---the core---is inaccessible in laboratory experiments, thus astrophysical insights are invaluable to our knowledge of strongly interacting superdense matter \cite{LattimerAnnRevNucPhys2021,Chatziioannou+ArXiv2024,TheChapter}.  

The EoS is linked to neutron star observables through the solution of the hydrostatic equilibrium equations. For a non-rotating static star (which is a good approximation even for most of millisecond pulsars), this problem is described by the Tolman-Oppenheimer-Volkoff (TOV) equations \cite{Tolman1939,OppVol1939} with a given initial condition in the star center (i.e., fixed central density or pressure). Solving TOV equations for a set of initial conditions, one can map the pressure $P$ -- energy density $\varepsilon = \rho c^2$ relation ($c$ is the vacuum speed of light) to the mass $M$ -- radius $R$ curve. This mapping is called the Oppenheimer-Volkoff mapping (OVM) \cite{Lind1992}. Other bulk observables like moment of inertia and tidal deformability rely on hydrostatic NS configurations \cite{Chatziioannou+ArXiv2024} and can be well approximated as functions of $M$ and $R$ \cite{LattimerPrakashApJ2001,YagiYounesPRD2013,YagiYounesPRD2013,YagiYunesCQG2016}, thus we consider OVM as a mapping from $P-\rho$ dependence 
to mass dependencies of all such quantities. This mapping is a bijection \cite{GerlachPhysRev1968,Lind1992}; thus, there exists the inverse Oppenheimer-Volkoff mapping (IOVM) from $M-R$ relation back to the EoS.

Each NS EoS inference from observations involves IOVM at some stage, and there are numerous implementations of this procedure in the literature. The most common today is an implicit way combined with the Bayesian approach. One considers a large sample of possible EoSs, constructs a direct OVM for each, evaluates the observational likelihoods for the resulting $M-R$ curves, assigns these likelihoods to the corresponding initial EoS, and ultimately performs a Bayesian inference of the EoS using these likelihoods. This scheme is usually supplemented with some efficient sampling method, like various Markov Chain Monte Carlo (MCMC) \cite{Foreman-Mackey+PASA2013_emcee} or nested sampling \cite{Feroz+MNRAS_MULTINEST} algorithms. Many constraints on the NS EoS are obtained in this way \cite[e.g.][]{Nattila+AA2016,Annala+Nat2023,Brandes+PRD2023,Rutherford2024,BrandesWeisePRD2025,Koehn+PRX2025}.

A systematic bias due to a specific form of the parameterization chosen in this procedure is hard to take into account. In particular, it is still debatable whether the phase transition to deconfined quark matter occurs in existing NSs or happens at densities higher than reached in the most massive ones \cite{Brandes+PRD2023,KomoltsevPRD2024_FOPT,TangHuangFanPRD2025,BrandesWeisePRD2025,EckerJokelaJarvinenArXiv2025}. To eliminate this kind of systematics, different authors use non-parametric approaches \cite[e.g.][]{CuceuRoblesPRD2025,BiswasArXiv2025} or try to find the ``best'' way to parameterize the EoS. It is always debatable whether the optimal parameterization should be purely phenomenological or rely on some set of EoS models derived from nuclear physics \cite[see, e.g.,][]{Lindblom2010,BrandesWeiseJCAP2024}. The first option is more flexible and can catch something beyond our up-to-date theoretical knowledge on superdense matter, but suffers from high number of potentially non-physical degrees of freedom. Vice versa, the second option aims to keep only relevant parameters, but makes the method valid only if the true EoS is somewhere amidst the existing set of models. 

An explicit way to perform IOVM is also presented in the literature \cite[e.g.][]{Lind1992,Soma+JCAP2022,BrandesWeiseJCAP2024,OfPirShtPRD2024,SunLattimerApJ2025}. Among numerous techniques, including artificial neural networks, we would like to emphasize the meta-model approach, which relies on universal properties of a (supposedly comprehensive) set of EoS models \cite{SunLattimerApJ2025,OfPirShtPRD2024}. In this work, we will follow the approach of our previous paper \cite{OfPirShtPRD2024}, hereafter \paperi, where we find that a wide and diverse manifold of models for EoS in the core can be described by a universal three-parametric approximation. The advantage of this approach is that it uses the most compact EoS parameterization that is motivated by state-of-the-art nuclear physics. Of course, it comes together with loss of flexibility: this method does not work if the true EoS is far from the manifold of existing theoretical models. Analysis of the current inferences on the NS EoS \cite{Rutherford2024,BrandesWeisePRD2025,Koehn+PRX2025,OfPirShtPRD2024} shows that this is not the case: credibility bands for the $P-\rho$ relation reside within the range spanned by theoretical models. The speed of sound behavior is more complicated, but the method of \paperi\ is calibrated for inferring only $P-\rho$ relation, not its derivative. 
{Applying this method to existing theoretical and observational constraints on NS properties, we set limits on the NS EoS. They are particularly stringent at the high-density end of the EoS.}

This work consists of three parts. First, we update the results
for the EoS obtained in
\paperi, including new observations and implementing theoretical constraints that were not addressed there (chiral effective field theory at low densities \cite{Keller+PRL2023}, causal extrapolation of perturbative quantum chromodynamics at high densities \cite{KomoltsevKurkelaPRL2022}; Sec.~\ref{sec:basic}). Second, we examine the impact of different observations on the total EoS inference (Sec.~\ref {sec:Roff-Moff}). We show that all existing observations that constrain NS radii improve the knowledge of the EoS quantitatively, but not qualitatively. This result is consistent with recent studies by Koehn~et~al.~\cite{Koehn+PRX2025}, who used an implicit IOVM with phenomenological speed-of-sound parameterization. Third, we investigate how future NS radius measurements can improve the EoS inference (Secs.~\ref{sec:future} and~\ref{sec:concl}). We show that to improve the existing limits on the EoS effectively, the observation (or its interpretation) has to tighten the constraints on $\Mmax$.

\section{Up-to-date EoS inference}
\label{sec:basic}

\subsection{Inverse Oppenheimer-Volkoff mapping for 3-parametric EoSs}
\label{sec:basic:IOVM}

We use the IOVM method developed in our previous works \cite{Ofengeim2020,OfShtPirAstLett2023,OfPirShtPRD2024}. It is based on the idea that the Oppenheimer-Volkoff limits for NS characteristics (i.e. properties of the maximum-mass NS) set universal scales for corresponding properties of a NS with arbitrary mass \cite{Ofengeim2020,Cai+ApJ2023}. We denote mass and radius of the maximum-mass NS as $\Mmax$ and $\Rmax$, and pressure, density, and the speed of sound in the center of this star as $\Pmax$, $\rhomax$, and $\csmax$. For a wide range of EoS models, these five  quantities are tightly correlated, so one can consider only two of them, say $\Pmax$ and $\rhomax$, as independent variables \cite{Ofengeim2020,Cai+ApJ2023,OfShtPirAstLett2023}. In \paperi\ we showed that the whole EoS in the range%
\footnote{
We adopt the nuclear saturation density $\rho_0 = 2.8\times 10^{14}\,\text{g}\,\text{cm}^{-3}$, $\rho_0 c^2 = 2.5\times 10^{35}\,\text{dyn}\,\text{cm}^{-2} = 157\,\text{MeV}\,\text{fm}^{-3}$ as a convenient unit for density and pressure of superdense matter.
} $\rho_0...\rhomax$ can be approximated by a three-parametric function of density $P(\rho; \Pmax, \rhomax, \ratio)$, where $\ratio \equiv \Rhalf/\Rmax$ is a ratio of the radius $\Rhalf = R(\Mmax/2)$ and the radius of the maximum mass NS. The corresponding $M-R$ relation in the range $\Msun...\Mmax$
can be approximated by a function $R(M; \Pmax, \rhomax, \ratio)$ with the same three parameters. To infer the EoS from observational data on masses and radii (and thus to perform an IOVM), one has to fit 
the data with the $R(M; \Pmax, \rhomax, \ratio)$ in order to constrain the triad $\rhomax$, $\Pmax$, $\ratio$. These constraints determine the EoS by using the $P(\rho; \Pmax, \rhomax, \ratio)$ relation. 

In addition to the observed data we include several theoretical constraints. 
In \paperi\ we used the causality condition $\csmax(\Pmax,\rhomax) \leqslant c$ (up to some safety factor that accounts for approximation errors), and the phenomenological constraint $\ratio > 1$, as we could not find an EoS model in the literature that violates the latter. In this work, we include additional constraints and correspondingly improve the universal three-parametric approximations.

\subsection{Updates of the method and the current state of the EoS inference}
\label{sec:basic:updates}

We update the method from \paperi\ in several aspects. From the theoretical point of view, we include:
\begin{itemize} 
    
    \item \textit{Perturbative quantum chromodynamics (pQCD) constraints}. We use the upper limit on $\Pmax$ as a function of $\rhomax$ derived in Ref.~\cite{KomoltsevKurkelaPRL2022} (see their figure~4) from matching the pQCD limit at densities much greater than $\rhomax$ according to the causality principle. 
    Notice that this method for inclusion of the pQCD constraint does not rely on interpolation of the EoS at densities above $\rhomax$, so it is free of bias that may arise in other works \cite[e.g.][]{Annala+Nat2023,Brandes+PRD2023,Rutherford2024,Koehn+PRX2025}. Here, the pQCD limit affects the EoS inference most conservatively. 
    For the relevant range of EoS parameters, accounting for the pQCD limit makes a virtually negligible enhancement of the causality condition $\csmax \leqslant c$ used in \paperi.

    \item \textit{$\chi$EFT corridor} (also \textit{chiral corridor}). We use the results of  Ref.~\cite{Keller+PRL2023} for the low-density part of the EoS, $\rho < 1.5\rho_0$, as a prior for fitting the EoS from observations. We impose this constraint
    in a manner similar to that in Ref.~\cite{Rutherford2024}, approximating it as a uniform corridor for the $P-\rho$ dependence:
    \begin{equation}
        \label{eq:chicorridor}
        1.51\times 10^{-2} \left( \frac{\rho}{\rho_0}\right )^{2.34} < \frac{P(\rho)}{\rho_0 c^2} < 2.09\times 10^{-2} \left( \frac{\rho}{\rho_0} 
        \right)^{2.77}.
    \end{equation}
    In terms of the Bayesian approach, we set a uniform prior for pressure at $\rho = 1.5\rho_0$ within this corridor.

    \item \textit{Improved universal $P-\rho$ approximation}, optimizing it for the density range above the chiral corridor, $1.5\rho_0...\rhomax$, instead of the range $\rho>\rho_0$ used in \paperi. We give all formulae for the approximations involved in the method in Appendix~\ref{sec:app:approxes}.
    
\end{itemize}

\begin{table}
\centering
\caption{\label{tab:updates}
Comparison of theoretical and observational constraints on NS properties used in \paperi\ and in Sec.~\ref{sec:basic} of the present work. Corresponding EoS inferences are illustrated by Fig.~\ref{fig:PaperI-vs-basic}. We use specific references instead of checkmarks when we need to emphasize that the key difference is the data source; see \paperi\ for other references. For the observational data used in the \basic\ scenario, numbers and capital letters in the last column indicate how they are labeled in Fig.~\ref{fig:PaperI-vs-basic}.
}
\small
\begin{tabular}{l@{\extracolsep{10pt}}cccr@{\extracolsep{2pt}}l}
\hline\hline
 & {\it \paperi} & \makecell{\it upd\\ \it obs} & \makecell{\it upd\\ \it theory} & \multicolumn{2}{c}{\makecell{\it upd both\\ (\basic)}} \\
\hline
\multicolumn{6}{c}{\bf Theoretical constraints} \\
Causality                  & $\checkmark$ & $\checkmark$ & $\checkmark$ & $\checkmark$ & \\
$\chi$EFT \& upd $P(\rho)$ &              &              & $\checkmark$ & $\checkmark$ & \\
pQCD                       &              &              & $\checkmark$ & $\checkmark$ & \\
$r_{1/2} > 1$                & $\checkmark$ & $\checkmark$ & $\checkmark$ & $\checkmark$ & \\
\hline\hline
\multicolumn{6}{c}{\bf Observations} \\
\multicolumn{6}{c}{\it Masses of radio pulsars} \\
PSR J1614--2230            & \cite{Demorest+Nat2010_1614} & \cite{NANOGrav15years} & \cite{Demorest+Nat2010_1614} & \cite{NANOGrav15years} & \textsc{a} \\
PSR J0348+0432             & \cite{Antoniadis+Sci2013_0348}          &              & \cite{Antoniadis+Sci2013_0348}          &              & \\
\hline
\multicolumn{6}{c}{\it Masses of spiders} \\
PSR J1810--1744            & $\checkmark$ & $\checkmark$ & $\checkmark$ & $\checkmark$ & \textsc{b} \\
PSR J1653--0158            & $\checkmark$ & $\checkmark$ & $\checkmark$ & $\checkmark$ & \textsc{c} \\
PSR J1311--3430            & $\checkmark$ & $\checkmark$ & $\checkmark$ & $\checkmark$ & \textsc{d} \\
PSR J2215+5135             & $\checkmark$ &              & $\checkmark$ &              & \\
PSR J0952--0607            & \cite{Romani+ApJL2022_0952} & \cite{Romani+2026_0952} & \cite{Romani+ApJL2022_0952} & \cite{Romani+2026_0952} & \textsc{e} \\
\hline
\multicolumn{5}{c}{\it Electromagnetic counterparts of mergers} \\
\makecell[l]{AT2017gfo \\ \&\ GRB~170817A}                   & $\checkmark$ & $\checkmark$ & $\checkmark$ & $\checkmark$ & \textsc{kn}\footnote{
The EM afterglow of GW170817 (AT2017gfo and GRB~170817A) is more than just a kilonova emission. Nevertheless, we denote the upper limit on $\Mmax$ that follows from its interpretation~\cite{RezzollaMostWeihApJL2018,NathanailMostRezzollaApJL2021} as KN.
} \\
\hline
\multicolumn{6}{c}{\it Masses and radii from GW events} \\
GW170817                   & $\checkmark$ & $\checkmark$ & $\checkmark$ & $\checkmark$ & 1 \\
\hline
\multicolumn{6}{c}{\it Masses and radii from X-ray spectroscopy} \\
PSR J0740+6620             & \cite{Riley+ApJL2021} & \cite{Salmi+ApJ2024_0740} & \cite{Riley+ApJL2021} & \cite{Salmi+ApJ2024_0740} & 2 \\
PSR J0030+0451             & \cite{Miller+2019ApJ0030} & \cite{Vinciguerra+ApJ2024} & \cite{Miller+2019ApJ0030} & \cite{Vinciguerra+ApJ2024} & 3 \\
PSR J0437--4715            & $\checkmark$ & $\checkmark$ & $\checkmark$ & $\checkmark$ & 4 \\
PSR J0614--3329            &              & $\checkmark$ &              & $\checkmark$ & 5 \\
PSR J1231--1411            &              & $\checkmark$ &              & $\checkmark$ & 6 \\
Cas A                      & $\checkmark$ & $\checkmark$ & $\checkmark$ & $\checkmark$ & 7 \\
4U J1702--429              & $\checkmark$ & $\checkmark$ & $\checkmark$ & $\checkmark$ & 8 \\
4U 1724--307               & $\checkmark$ & $\checkmark$ & $\checkmark$ & $\checkmark$ & 9 \\
SAX J1810.8--260           & $\checkmark$ & $\checkmark$ & $\checkmark$ & $\checkmark$ & 10 \\
\hline\hline
\end{tabular}
\end{table}

To the set of observational data, we perform the following modifications:
\begin{itemize}

    \item {\it Revised mass} $1.937\pm 0.014\,\Msun$ inferred for PSR~J1614--2230 in the last data release of the NANOGrav collaboration~\cite{NANOGrav15years}, instead of the original results $1.97\pm 0.04\,\Msun$ \cite{Demorest+Nat2010_1614}.

    \item {\it Tightened mass of PSR~J0952--0607} $2.35\pm 0.11\,\Msun$~\cite{Romani+2026_0952} versus the old result $2.35\pm 0.17\,\Msun$ \cite{Romani+ApJL2022_0952}.
    
    \item {\it Excluding PSR~J0348+0432} from consideration, since its mass is remeasured from $2.01\pm 0.04\,\Msun$ \cite{Antoniadis+Sci2013_0348} used in \paperi\ to lower values $1.806\pm 0.037\,\Msun$ \cite{Saffer+ApJL2025_0348}. As the radius of this star is unknown, it only constrains $\Mmax$ from below, and this constraint is negligible with respect to those coming from other heavier NSs.

    \item {\it Excluding  PSR~J2215+5135}, since the recent study of its mass~\cite{SullivanRomaniApJ2024_2215} shows that the results are not robust with respect to model systematics.
    
    \item {\it Updated data} for PSR~J0740+6620 and PSR~J0030+0451 according to the ``new'' scenario of Ref.~\cite{Rutherford2024}, i.e. we use the results of Ref.~\cite{Salmi+ApJ2024_0740} instead of Ref.~\cite{Riley+ApJL2021} for the former and the output of the ST+PDT run from Ref.~\cite{Vinciguerra+ApJ2024} instead of Ref.~\cite{Miller+2019ApJ0030} for the latter. 
    
    \item {\it Two new NICER objects}, PSR~J0614--3329 \cite{Mauviard+2025_0614} (their `Headline' scenario) and PSR~J1231--1411 \cite{Salmi+ApJ2024_1231}(`PDT-U H $R_\text{eq} \in [10,\,14]\,$km' scenario). Results for the latter NS are difficult to interpret: they tend to anomalously large radii and masses anomalously small, with the posterior artificially truncated at $M < 1\,\Msun$ and $R > 14\,$km. However, among the two runs that are independent of the EoS inference from previous works of the same group, this is the only useful one since for the other one the Monte Carlo fitting procedure has not converged~\cite{Salmi+ApJ2024_1231}.
    
\end{itemize}

The NANOGrav collaboration reported several Shapiro-delay-derived measurements of NS masses in addition to PSR~J1614--2230 \cite{NANOGrav15years}. Among them, the most important are those with a significant fraction of their mass posterior at $M > 2\,\Msun$. Unfortunately, most of those estimates suffer from low statistics. Thus, we do not include this data in the analysis performed in the main text. We study its impact in Appendix~\ref{sec:app:NNG15}.

The rest of the observations used here are the same as in \paperi, and we refer the reader there for corresponding references. In Table~\ref{tab:updates}, we compile theoretical constraints and observational data used in our analysis. We group the data into four sets: 
{\it \paperi:} similar to \paperi; 
{\it upd obs:} updated observations but ``old'' theoretical aspects; 
{\it upd theory:} ``old'' observations but updated theoretical methods;
{\it upd both}, or \basic: both theory and observations updated. 
The latter set is hereafter addressed as the basic scenario for fitting the NS EoS. Notice that in all scenarios we use only a single gravitational wave (GW) event GW~170817 \cite{GW170817}.  We do not use other GW events  with a claimed NS component \cite[e.g.][]{GW190425_2020ApJ,GW190814_2020ApJ}, as their impact on the EoS inference is negligible.

\begin{figure*}
    \centering
    \includegraphics[width=0.99\textwidth]{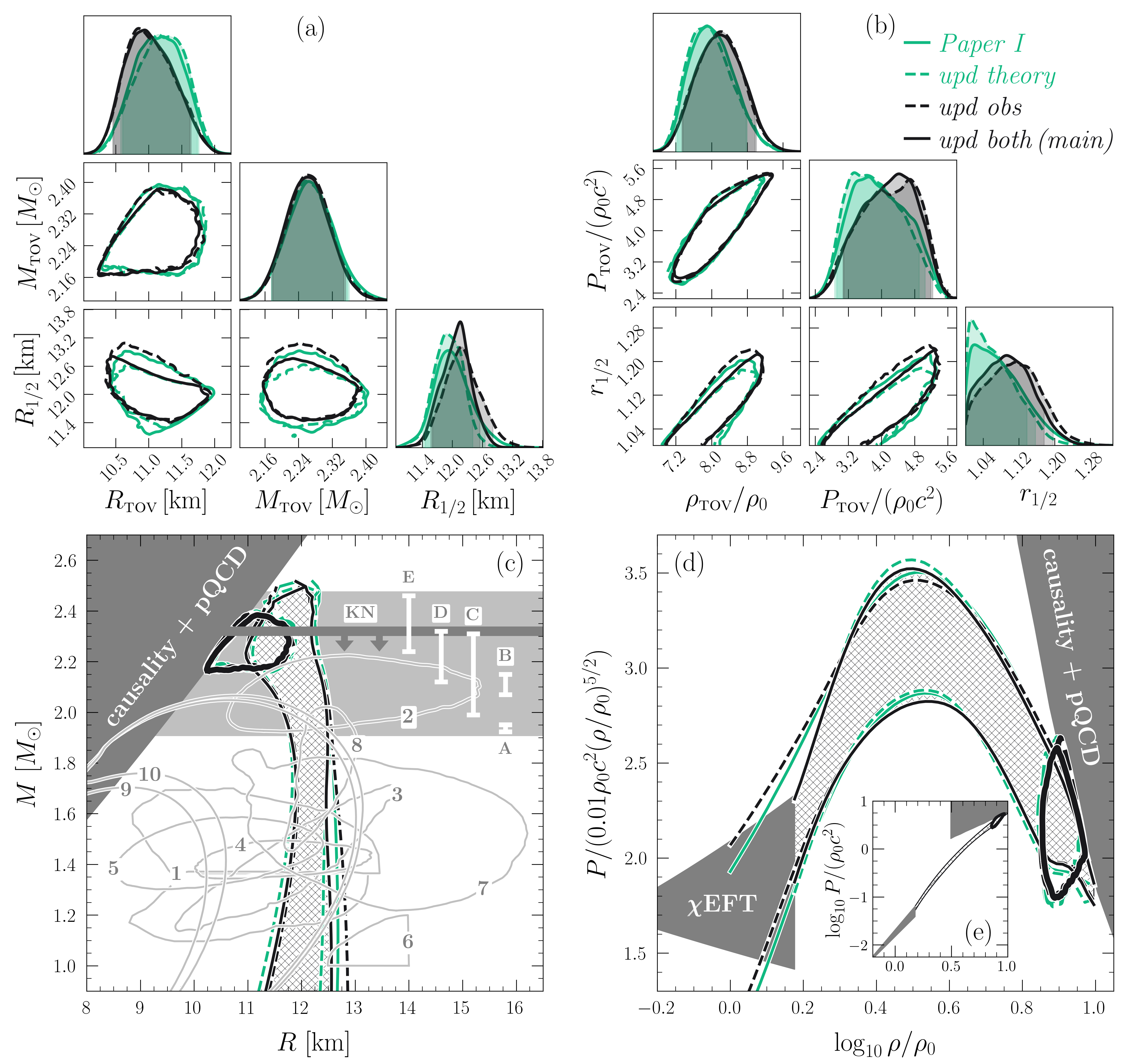}
    \caption{A comparison of the EoS constraints obtained in \paperi\ with those obtained with updated theoretical and/or observational constraints. 
    (a) Pairwise correlations and one-dimensional marginalized posterior distributions for the parameters of the universal $M-R$ curve approximation: maximum NS mass $\Mmax$, radius of such a star $\Rmax$, and radius $\Rhalf$ of a NS with $M=\Mmax/2$. 
    (b) Pairwise correlations and one-dimensional marginalized posteriors for the parameters of the universal EoS approximation: density $\rhomax$ and pressure $\Pmax$ in the center of the maximum-mass NS, and the ratio $\ratio = \Rhalf/\Rmax$. 
    (c) Posterior credibility bands for the $M-R$ curve (thin black or green lines) compared to NS observations. The \basic\ scenario is highlighted by cross-hatch. Thick contours show the credible areas for the $(\Mmax,\Rmax)$ point. Number-labeled gray contours show credible regions for NSs with both $M$ and $R$ measured. Letter-labeled vertical bars display 1$\sigma$ mass ranges for NSs in binaries. See the labels description in Table~\ref{tab:updates}. The thick horizontal gray strip shows the overall range of their masses. The thick horizontal line shows the upper limit on $\Mmax$ adopted from the GW 170817 electromagnetic counterpart interpretation. 
    (d) Posterior credibility bands for the EoS. The main trend $P\propto \rho^{2.5}$ is factored out . The \basic\ scenario is highlighted by a cross-hatch. Thick contours show the credible areas of the $(\Pmax,\rhomax)$ point. The gray-shaded ``$\chi$EFT'' area shows the chiral corridor Eq.~\eqref{eq:chicorridor}. The ``causality+pQCD'' region is forbidden according to \cite{KomoltsevKurkelaPRL2022}. (e) The same as panel (d) for the \basic\ scenario, but full pressure is shown on the vertical axis (log-scaled). All credibility areas are 90\%.}
    \label{fig:PaperI-vs-basic}
\end{figure*}

These four sets are comprehensively analyzed in Fig.~\ref{fig:PaperI-vs-basic}, which presents our ``standard protocol'' for the NS EoS inference within the three-parametric approach. The top row, Figs.~\ref{fig:PaperI-vs-basic}a and~\ref{fig:PaperI-vs-basic}b, shows corner plots for three EoS parameters. The bottom row, Figs.~\ref{fig:PaperI-vs-basic}c and~\ref{fig:PaperI-vs-basic}d, demonstrates the EoS behavior on the $M-R$ and $P-\rho$ planes. Two left panels \ref{fig:PaperI-vs-basic}a and \ref{fig:PaperI-vs-basic}c show the astrophysical representation of the EoS in terms of masses and radii. The right column displays the EoS in terms of pressure and density. The triad $(\Mmax,\Rmax,\Rhalf)$, Fig.~\ref{fig:PaperI-vs-basic}a, conveniently parameterize the $M-R$ curve, see Eqs.~\eqref{eq:MRfit} and Paper~I. The triad $(\Pmax,\rhomax,\ratio)$, Fig.~\ref{fig:PaperI-vs-basic}b, is suitable for parameterization of the $P-\rho$ dependence, see Eqs.~\eqref{eq:Prhofit}. OVM bijectivity establishes one-to-one correspondence between $M-R$ curves in Fig.~\ref{fig:PaperI-vs-basic}c and $P-\rho$ dependencies in Fig.~\ref{fig:PaperI-vs-basic}d, as well as between the triads $(\Mmax,\Rmax,\Rhalf)$ and $(\Pmax,\rhomax,\ratio)$, see Eqs.~\eqref{eq:maxCorrs}. Notice that on the vertical axis of the $P-\rho$ plane, Fig.~\ref{fig:PaperI-vs-basic}d, we display the pressure normalized to the average trend of all NS EoSs, empirically determined as $P \propto \rho^{5/2}$. 

Fig.~\ref{fig:PaperI-vs-basic}c illustrates a subtle but expected feature~\cite{OfPirShtPRD2024} of the inference: the two-dimensional credibility contours for the position of the $(\Rmax,\Mmax)$ point are systematically shifted to the left with respect to the top end of the credibility area for the whole $M-R$ curve. The explanation follows from the definition of the $M-R$ credibility area as the stacking of one-dimensional credible intervals of $R$ at each fixed mass. Since each $M-R$ curve in the 
posterior sample
behaves almost horizontally near its maximum point, the values of $R(M)$ at fixed $M$ in the middle of $\Mmax$ range are systematically larger than the values of $\Rmax=R(\Mmax)$.

Three theoretical constraints, pQCD at high densities, $\chi$EFT at low densities, and $\ratio>1$, also mainly affecting low $\rho$, are clearly visible in the right column of Fig.~\ref{fig:PaperI-vs-basic}. In the two left panels, the latter two constraints are less explicit, while the pQCD limit is mapped to the $M-R$ plane via the $\Mmax,\Rmax - \Pmax,\rhomax$ correlations Eqs.~\eqref{eq:maxCorrs}. 

\begin{table*}
\centering
\caption{\label{tab:ns_parameters} EoS parameters for the \basic\ scenario, the highest probability value and 95\% uncertainties.}
\small
\renewcommand{\arraystretch}{1.4}
\setlength{\tabcolsep}{5pt}
\begin{tabular}{ccc|ccc|ccc}
\hline\hline
$\Rmax$ [km] & $\Mmax$ [$\Msun$] & $\Rhalf$ [km] & $\rhomax/\rho_0$ & $\Pmax/(\rho_0 c^2)$ & $\ratio$ & $R_{1.4}$ [km] & $R_{2.0}$ [km] & $R_{2.0}-R_{1.4}$ [km] \\
\hline
  $10.92_{-0.54}^{+0.79}$ 
& $2.26_{-0.10}^{+0.10}$ 
& $12.15_{-0.66}^{+0.49}$ 
& $8.18_{-0.95}^{+0.94}$ 
& $4.43_{-1.50}^{+0.86}$ 
& $1.09_{-0.09}^{+0.09}$ 
& $12.30_{-0.55}^{+0.26}$ 
& $12.06_{-0.47}^{+0.50}$ 
& $-0.09_{-0.53}^{+0.50}$ \\
\hline\hline
\end{tabular}
\end{table*}

Overall view of Fig.~\ref{fig:PaperI-vs-basic} indicates that the updates cause only minor changes of the inference yielded in \paperi. This includes both updated data and updated theoretical constraints. 
Taken separately, these updates affect the EoS inference in opposite directions. This results in tighter limits on different parameters.  
The most prominent change occurs at low densities (and masses) due to the implementation of the chiral corridor. The chiral corridor that is imposed here  pushes the upper boundary of the EoS credibility band down. One can see this effect by comparing the solid green \textit{\paperi} and dashed green \textit{upd theory} lines in the $M-R$ and $P-\rho$ planes. Comparison of dashed black \textit{upd obs} and solid black \basic\ in Fig.~\ref{fig:PaperI-vs-basic}d (i.e. both theory and observations updated) shows the same systematic shift as compared to the results of \paperi.  As for one-dimensional distributions of the EoS parameters shown in Figs.~\ref{fig:PaperI-vs-basic}a,b, the main change comes from updating observational data, not the theoretical constraints. Remarkably,  the $\Mmax$ distribution remains virtually unchanged.

However, all these changes between \paperi\ and the \basic\ scenario of this work are subdominant and have low statistical significance. The main outcome of the comparison made in Fig.~\ref{fig:PaperI-vs-basic} is that the old and updated sets of constraints on NS properties give well-consistent EoS inferences. 

\begin{figure}
    \centering
    \includegraphics[width=\columnwidth]{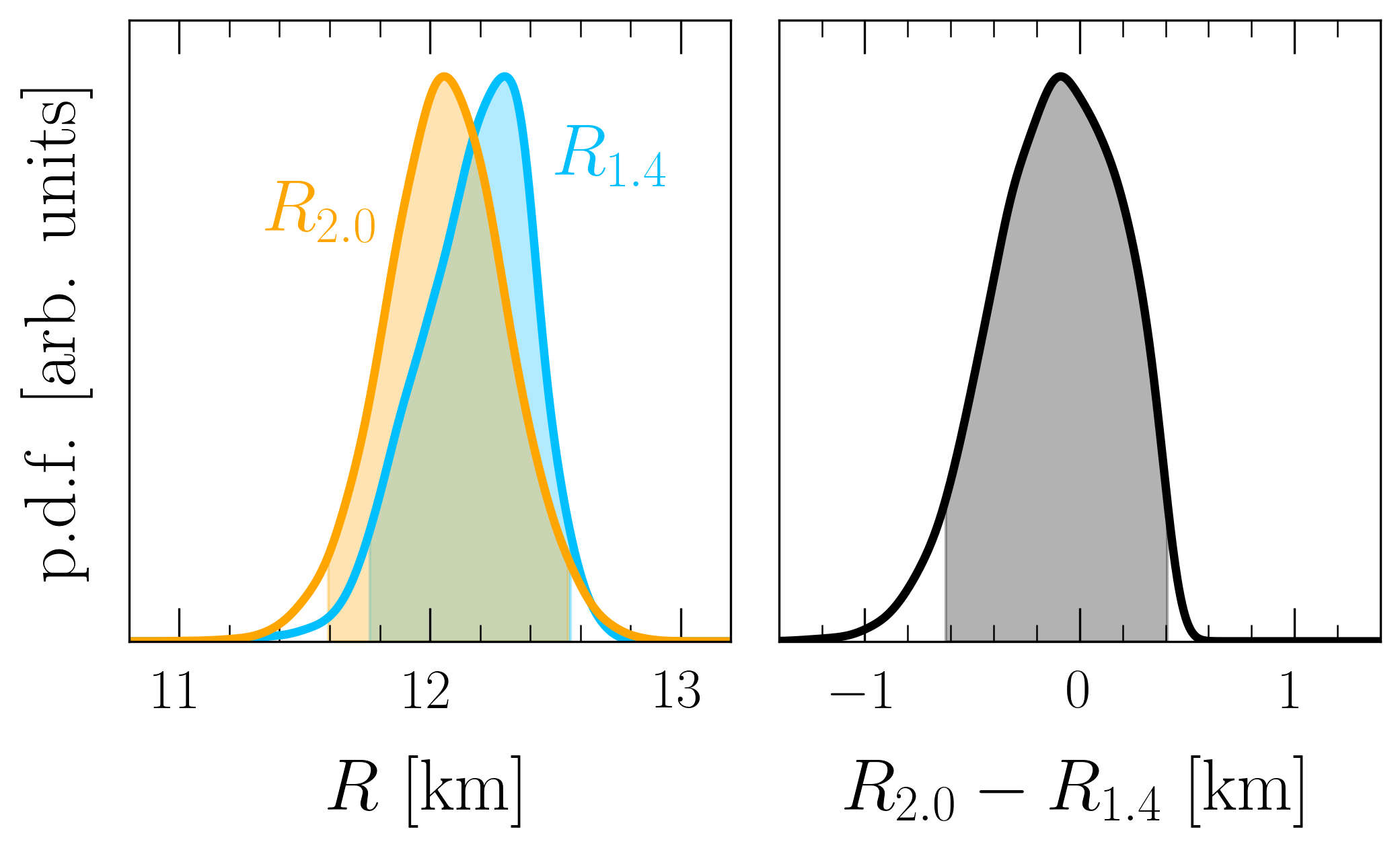}
    \caption{{Marginalized posterior distributions} for the radius of the canonical NS $R_{1.4}$, the radius of $2\,\Msun$ NS $R_{2.0}$, and $ R_{2.0}-R_{1.4}$ (\basic\ data set). Shaded areas show 95\% credibility.}
    \label{fig:R14R20}
\end{figure}

In Table~\ref{tab:ns_parameters}, we list the results for the key EoS parameters under the \basic\ scenario. We give the astrophysical representation $\Mmax$, $\Rmax$, $\Rhalf$, the $P-\rho$ relation representation $\rhomax$, $\Pmax$, $\ratio$, and additionally the parameters that are often considered as important EoS markers in the astrophysical literature \cite{Drischler+PRC2021,LattimerAnnRevNucPhys2021,Rutherford2024,Mauviard+2025_0614}: the radius $R_{1.4}$ of the canonical NS with $M = 1.4\,\Msun$, the radius $R_{2.0}$ of $2\,\Msun$ NS, and the difference $ R_{2.0}-R_{1.4}$ (see also Fig.~\ref{fig:R14R20}). The latter quantity is an indicator of the EoS stiffness: positive for stiff EoS and negative for soft EoS \cite{Drischler+PRC2021}. Our result favors moderate stiffness with $R_{1.4} \approx R_{2.0}$ and disfavors both strongly negative and positive $R_{2.0}-R_{1.4}$ values.

\section{Relative impact of different observations}
\label{sec:Roff-Moff}

We turn now to explore which parts of these constraints---old and new, theoretical and observational---dominate the EoS inference.

\begin{table}
\centering
\caption{\label{tab:Moff-Roff}
A comparison of the \basic\ set of constraints 
and various partial data sets used in Sec.~\ref{sec:Roff-Moff}. We use specific references instead of checkmarks when we need to emphasize that the key difference is the data source. For the observational data used in the \basic\ scenario, capital letters and numbers indicate how they are noted in the figures.
}
\small
\begin{tabular}{l@{\extracolsep{10pt}}r@{\extracolsep{2pt}}lccc}
\hline\hline
 & \multicolumn{2}{c}{\basic} & \makecell{\it min+\\ \it masses} & \makecell{\it min+\\ \it radii} & \makecell{\mimimi} \\
\hline
\multicolumn{6}{c}{\bf Theoretical constraints} \\
Causality                  & $\checkmark$ &      & $\checkmark$ & $\checkmark$ & $\checkmark$ \\
$\chi$EFT \& upd $P(\rho)$  & $\checkmark$ &     & $\checkmark$ & $\checkmark$ & $\checkmark$ \\
pQCD                       & $\checkmark$ &      & $\checkmark$ & $\checkmark$ & $\checkmark$ \\
$r_{1/2} > 1$       & $\checkmark$ &      & $\checkmark$ & $\checkmark$ & $\checkmark$ \\
\hline\hline
\multicolumn{6}{c}{\bf Observations} \\
\multicolumn{6}{c}{\it Masses of radio pulsars} \\
PSR J1614--2230            & $\checkmark$ & \textsc{a} & $\checkmark$ & $\checkmark$ & $\checkmark$ \\
\hline
\multicolumn{6}{c}{\it Masses of spiders} \\
PSR J1810--1740            & $\checkmark$ & \textsc{b} & $\checkmark$ &              &              \\
PSR J1653--0158            & $\checkmark$ & \textsc{c} & $\checkmark$ &              &              \\
PSR J1311--3430            & $\checkmark$ & \textsc{d} & $\checkmark$ &              &              \\
PSR J0952--0607            & $\checkmark$ & \textsc{e} & $\checkmark$ &              &              \\
\hline
\multicolumn{6}{c}{\it Electromagnetic counterparts of mergers} \\
\makecell[l]{AT2017gfo \\ \&\ GRB~170817A}  & $\checkmark$ & \textsc{kn} & $\checkmark$ &             &              \\
\hline
\multicolumn{6}{c}{\it Masses and radii from GW events} \\
GW170817                   & $\checkmark$ & 1           &              & $\checkmark$ &               \\
\hline
\multicolumn{6}{c}{\it Masses and radii from X-ray spectroscopy} \\
PSR J0740+6620             & \cite{Salmi+ApJ2024_0740}  & 2           & \makecell[c]{$M$ only\\ from \cite{NANOGrav15years}} & \cite{Salmi+ApJ2024_0740} & \makecell[c]{$M$ only\\ from \cite{NANOGrav15years}} \\
PSR J0030+0451             & $\checkmark$  & 3           &              & $\checkmark$  \\
PSR J0437--4715            & $\checkmark$ & 4           &              & $\checkmark$ &             \\
PSR J0614--3329            & $\checkmark$ & 5           &              & $\checkmark$ &             \\
PSR J1231--1411            & $\checkmark$ & 6           &              & $\checkmark$ &             \\
Cas A                      & $\checkmark$ & 7           &              & $\checkmark$ &             \\
4U 1702--429              & $\checkmark$ & 8           &              & $\checkmark$ &              \\
4U 1724--307               & $\checkmark$ & 9           &              & $\checkmark$ &             \\
SAX J1810.8--2609          & $\checkmark$ & 10          &              & $\checkmark$ &             \\
\hline\hline
\end{tabular}
\end{table}

We divide the observational data into two groups: the data that constrain both NS radii and masses, and the data that constrain NS masses only. The first group contains the GW event and the stars for which X-ray spectroscopy is available. The second group includes high-mass pulsars
\footnote{
NSs with estimated masses, but without radius estimates constrain the EoS  predominantly via limiting $\Mmax$ (see the formalism of \paperi). Only the highest measured NS masses are important for this.
} (sources A--E in Fig.~\ref{fig:PaperI-vs-basic} and Table~ \ref{tab:updates})
and the upper limit on $\Mmax$ from the EM counterpart of GW170817 (KN in Fig.~\ref{fig:PaperI-vs-basic} and Table~ \ref{tab:updates}). 
We investigate the impact of these groups by including them sequentially into
the EoS fitting procedure and comparing the result with the \basic\ scenario. Thus, we define the following data sets:
\begin{itemize}

    \item The \mimimi\ data set includes only the minimal astrophysical data that ensures the most basic observational knowledge on the EoS, i.e. $\Mmax \gtrsim 2\,\Msun$. Technically, we keep the data on masses of PSRs J0740+6620 and J1614--2230, and discard the rest. In this case, for PSR~J0740+6620 we use the NANOGrav collaboration result \cite{NANOGrav15years} $M = 1.99\pm 0.07\,\Msun$ as the most conservative one.
    
    \item The \mon\ set includes all high-mass pulsars used in the \basic\ set (A--E in Fig.~\ref{fig:PaperI-vs-basic} and Table~\ref{tab:updates})
    and the upper limit from the EM counterpart of GW170817 (KN in Fig.~\ref{fig:PaperI-vs-basic} and Table~\ref{tab:updates}). The mass of PSR~J0740+6620 is again taken from Ref.~\cite{NANOGrav15years}. All NS radii measurements are discarded.
    
    \item The \ron\ set includes all objects from the \basic\ scenario with both $M$ and $R$ measured, but does not include mass estimates of the spider pulsars and the GW170817 EM counterpart. The radio pulsar J1614--2230 is included for the same reasons as in the \mimimi\ data set.
    
    \item Merging \mon\ and \ron\ sets, one gets the \basic\ scenario.
    
\end{itemize}

\begin{figure*}
    \centering
    \includegraphics[width=0.99\textwidth]{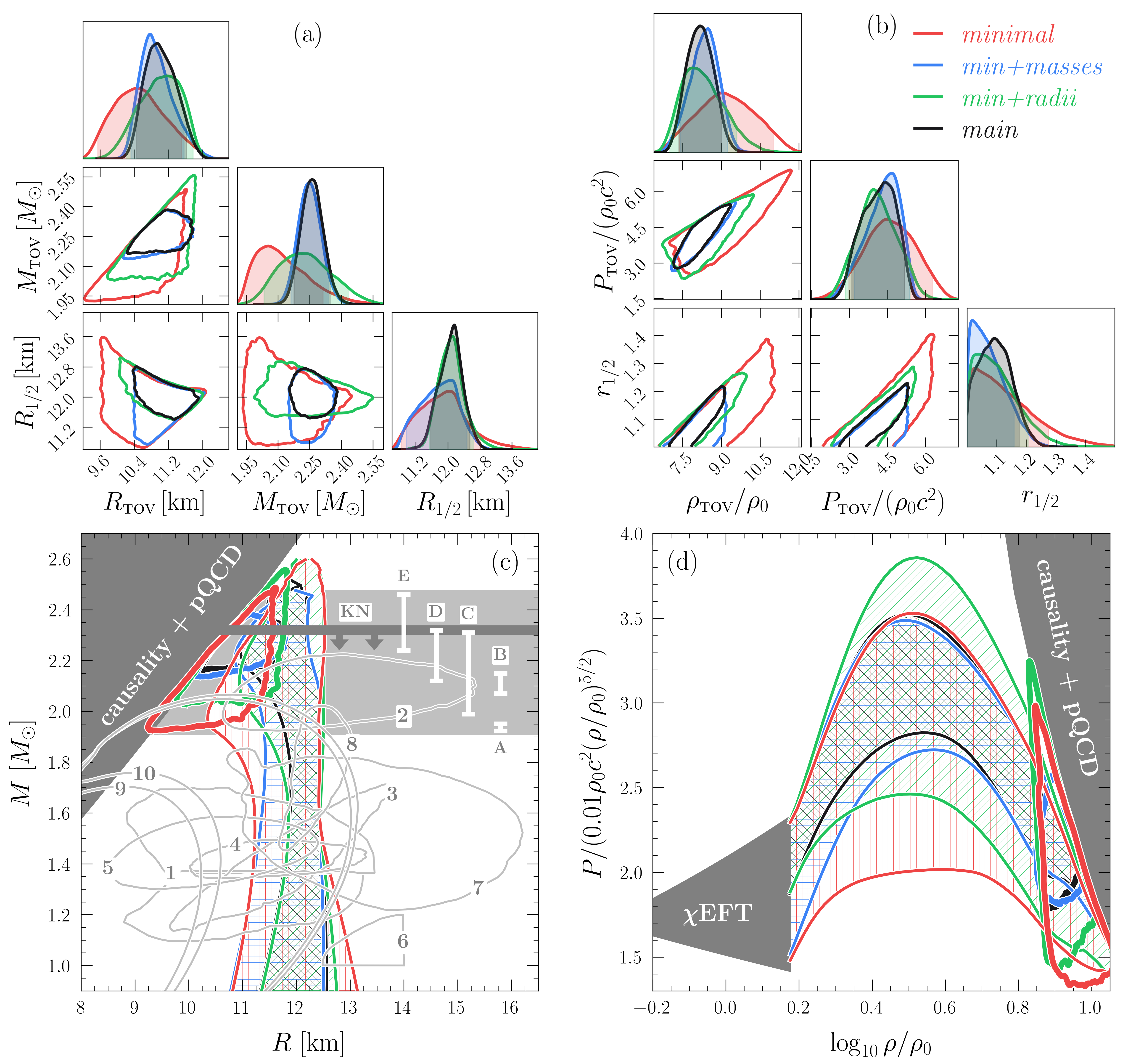}
    \caption{The same as Fig.~\ref{fig:PaperI-vs-basic} but for comparing the \basic\ scenario (shown in black) with \mon\ (blue), \ron\ (green), and \mimimi\ (red) scenarios, see Table~\ref{tab:Moff-Roff}.}
    \label{fig:Roff-Moff_main}
\end{figure*}

Summaries of constraints---theoretical and observational---used in these scenarios are given in the corresponding columns in Table~\ref{tab:Moff-Roff}. Fig.~\ref{fig:Roff-Moff_main} shows a  comparison of inferences on the EoS implied by the \basic, \mon, \ron, and \mimimi\ scenarios in the same way as in Fig.~\ref{fig:PaperI-vs-basic}. Shown in red is the \mimimi\ one, which demonstrates the widest range of EoSs that are compatible with the theoretical constraints and a very conservative lower limit $\Mmax \gtrsim 2\,\Msun$. In this case, the radii of low-mass NSs and the pressure at low densities are limited only by the $\chi$EFT corridor. In the $M-R$ plane, it makes the credible band for small and moderate masses, $M\lesssim 1.7\,\Msun$, twice wider (mainly extending to smaller radii) than in the \basic\ scenario. The two-dimensional distribution of $\Mmax$ and $\Rmax$ is limited by three constraints (and thus it has a triangular shape): causality + pQCD from the left top corner, $\Mmax$ greater than masses of the two massive pulsars, and the combination of $\Rmax < \Rhalf$ and the chiral corridor that constrains $\Rhalf$. In the $P-\rho$ plane, the $\Pmax-\rhomax$ distribution also resembles a triangle obtained from the $\Mmax-\Rmax$ distribution by the correlations between characteristics of the maximum mass NS, Eqs.~\eqref{eq:maxCorrs}. The overall width of the EoS credible band is 2--2.5 times wider than in the \basic\ scenario, primarily extending to lower pressures.

\begin{figure*}
    \centering
    \includegraphics[width=0.99\textwidth]{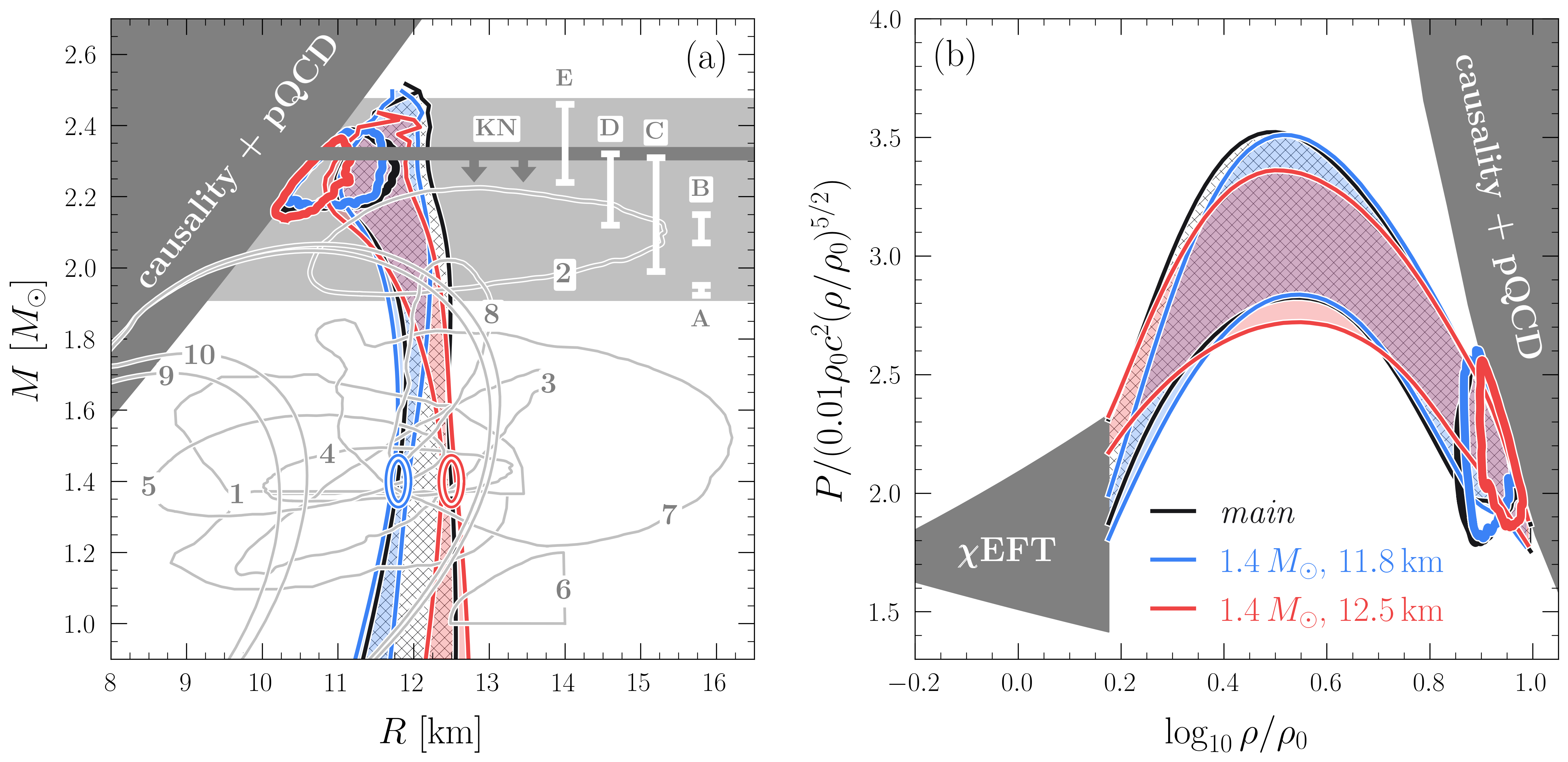}
    \caption{The same as panels (c) and (d) of Fig.~\ref{fig:PaperI-vs-basic} but for comparing the \basic\ scenario with the impact of a hypothetical observation of a $1.40\pm 0.05\,\Msun$ NS with $R=11.8\pm 0.1\,$km (blue) and $R=12.5\pm 0.1\,$km (red). All uncertainties are at 90\% level. The 90\% credible contours for masses and radii of these hypothetical stars are shown by double lines of the corresponding color in the panel (a).}
    \label{fig:Rtest_M=14}
\end{figure*}

Adding the data that directly tighten the $\Mmax$ constraints, we get the \mon\ data set, which is shown in blue in Fig.~\ref{fig:Roff-Moff_main}. The added constraints cut the upper and lower segments of the $\Mmax-\Rmax$ distribution, making it virtually coinciding with that for the \basic\ scenario. The correlations~\eqref{eq:maxCorrs} ensure that $\Pmax-\rhomax$ distributions for \mon\ and \basic\ sets also virtually coincide. In the $M-R$ plane (Fig.~\ref{fig:Roff-Moff_main}c), high-mass NS data strongly limits the high-mass end of the $M-R$ curve and keeps the low-mass end of it almost untouched. In the EoS plane (Fig.~\ref{fig:Roff-Moff_main}d), adding the high-mass data to the \mimimi\ set improves the inference on the $P-\rho$ relation for all densities but the lowest ones.

The green color in Fig.~\ref{fig:Roff-Moff_main} depicts the \ron\ scenario. Oppositely to the \mon\ scenario, it strongly limits the low-mass end of the $M-R$ curve, making it virtually coinciding with the \basic\ scenario for $M \lesssim 1.7\,\Msun$. Actually, it is the astrophysical data on NS radii that defines the lower boundary for the size of low-mass NSs. At high masses, \ron\ data set puts much looser constraints on the $M-R$ curve than one gets in the \basic\ scenario. In terms of the EoS, \ron\ scenario limits the $P-\rho$ relation only at the lowest densities, keeping the rest density range virtually unaffected with respect to the \mimimi\ scenario.

The \basic\ scenario in Fig.~\ref{fig:Roff-Moff_main} looks like an intersection of all other scenarios. Our findings are in agreement with the results by Koehn~et~al.~\cite{Koehn+PRX2025}, who have shown that posteriors of NS properties settle around ultimate results even when adding only a few constraints, relatively independent of the exact combination.

\section{Impact of possible future observations}
\label{sec:future}

\begin{figure*}
    \centering
    \includegraphics[width=0.99\textwidth]{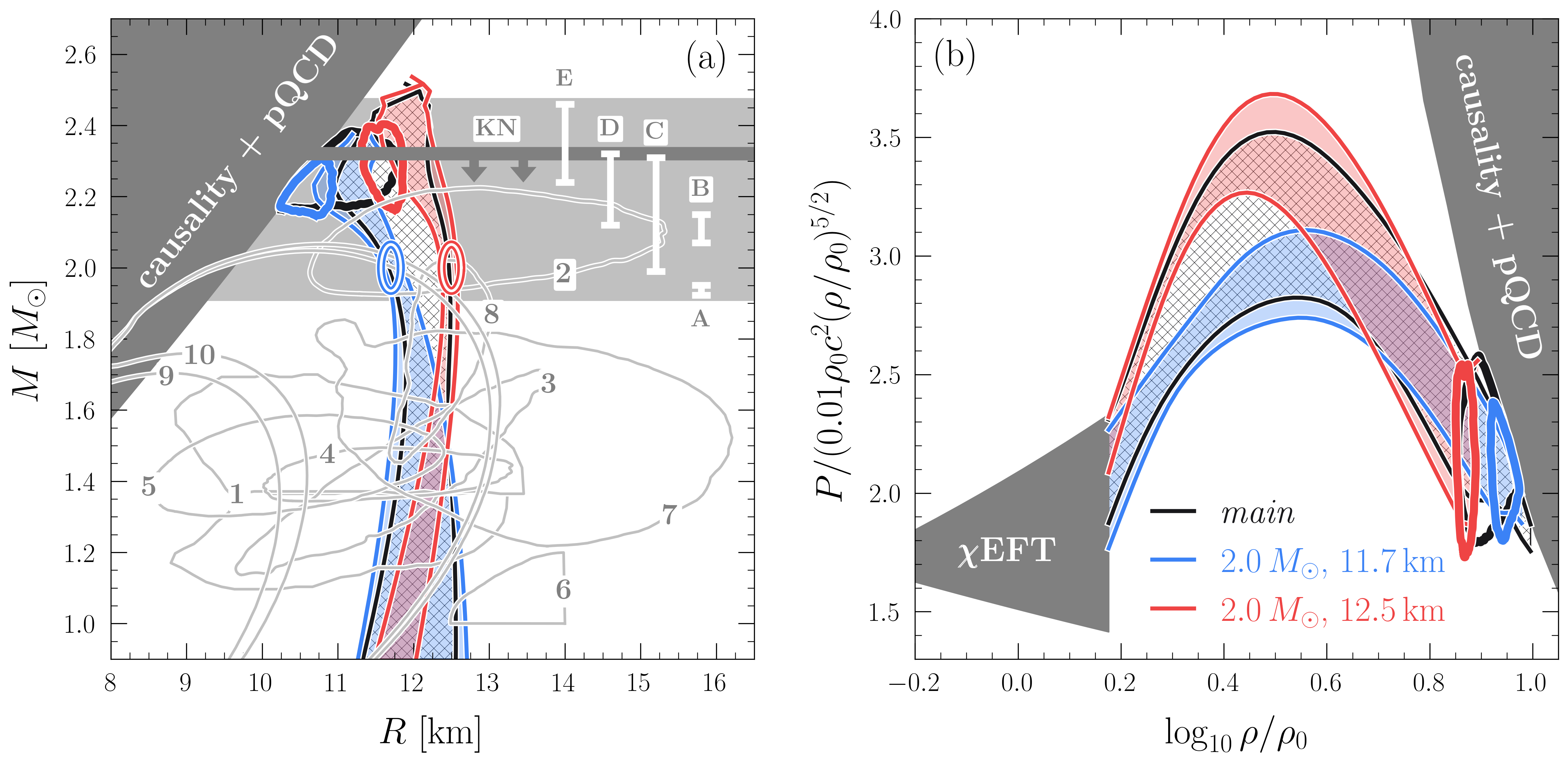}
    \caption{The same as Fig.~\ref{fig:Rtest_M=14} but for hypothetical observations of a $2.00\pm 0.05\,\Msun$ NS with $R=11.7\pm 0.1\,$km (blue) and $R=12.5\pm 0.1\,$km (red). All uncertainties are at 90\% level. The 90\% credible contours for masses and radii of these hypothetical stars are shown by double lines of the corresponding color in the panel (a).}
    \label{fig:Rtest_M=20}
\end{figure*}

We now turn to investigate the extent to which future NS observations may impact our understanding of the EoS. Obviously, different observations would have a different impact. For instance, one can speculate following the results of Sec.~\ref{sec:Roff-Moff} that a discovery of a new spider pulsar with a mass significantly larger than $2\,\Msun$ will likely be more useful than a measurement of mass and radius of a canonical NS ($M \sim 1.4\,\Msun$). In this Section, we qualitatively investigate the possible impact of such hypothetical future observations. 

\subsection{Hypothetical $M$ and $R$ observation}
\label{sec:future:m+r}

Suppose that we observe an NS with precisely measured mass and radius. To be specific, we assume 90\% uncertainties to be $0.05\,\Msun$ and $0.1\,$km. While for the mass, such accuracy is achievable today, current measurements of NS radii are significantly less accurate. Of course, the importance of such a single observation depends on the specific values of $M$ and $R$ observed. The greatest impact will be achieved if this observation strongly conflicts with the current inference on the EoS. For instance, observation of a star with $R\lesssim 10\,$km will provide powerful evidence for an extremely strong phase transition that creates a third generation of compact stars \cite[e.g.][]{Glendenning}. We keep consideration of such an option for our future research and refer the reader to, e.g., Refs.~\cite{Ayriyan+2025,KomoltsevPRD2024_FOPT,EckerJokelaJarvinenArXiv2025} for state-of-the-art investigations of the potential first-order phase transition in NS cores. In this work, we only consider hypothetical observations that are at least marginally consistent with the results of the \basic\ scenario considered above (Sec.~\ref{sec:basic:updates}). 

First, we consider observations of NSs with mass $M = 1.4\pm 0.05\,\Msun$. We take two options for its radius, $R = 11.8\pm 0.1\,$km and $R = 12.5\pm 0.1\,$km at 90\% credibility. Two-dimensional 90\% credibility contours for these cases are shown by blue and red double lines in Fig.~\ref{fig:Rtest_M=14}a. They represent two limiting cases of a canonical NS that is marginally consistent with the \basic\ set of observations. When we add such a hypothetical object to the \basic\ scenario, we obtain new 90\% credibility bands for the $M-R$ curve (Fig.~\ref{fig:Rtest_M=14}a) and EoS (Fig.~\ref{fig:Rtest_M=14}b). They are presented in Fig.~\ref{fig:Rtest_M=14} in blue and red colors, respectively. 

For low and moderate masses, the impact of such a hypothetical observation is valuable but not dramatic: the $M-R$ credibility band shrinks by a factor of $2-3$. The impact is much weaker for high masses. While it can make the $\Rmax$ posterior narrower (see the red case in Fig.~\ref{fig:Rtest_M=14}a), the $\Mmax$ posterior uncertainty remains the same. In the $P-\rho$ plane this observation improves the EoS credibility band only at $\rho \lesssim 2\rho_0$. This is expected as this lower range of the EoS dominates the radii of intermediate mass NSs. At higher densities, the effect of such an observation, even though it is significantly more accurate than current radius determinations, is not statistically significant. 

Second, we test the possible impact of an observation of an NS with $M = 2.0 \pm 0.04\,\Msun$. Again, we consider two possible values of radius that are marginally consistent with the \basic\ scenario, $R = 11.7\pm 0.1\,$km and $12.5\pm 0.1\,$km (90\%). Corresponding 90\% credibility contours are shown in Fig.~\ref{fig:Rtest_M=20}. In the $M-R$ plane, impact from such a hypothetical measurement essentially depends on the radius value: the NS with a large radius constrains the $M-R$ curve more strongly than the one with a small $R$. In the $P-\rho$ plane, this effect is less pronounced: the large-radius NS constrains the EoS at moderate densities stronger than the small-radius one, but at high densities the blue EoS credibility band (corresponding to small $R$) is even slightly narrower than the red band (for large $R$). This arises from the difference of $\Mmax$ constraints, which appear to be stronger in the small-$R$ case. Anyway, improvements of the EoS inference at $\rho \gtrsim 4\rho_0$ are not significant even for such a remarkably precise hypothetical observation.

\subsection{Hypothetical high $M$ observation}
\label{sec:future:m}

We turn now to investigate the impact of a hypothetical observation similar to that of a spider pulsar J0952--0607 or the radio pulsar J1614--2230, i.e. a measurement of a high NS mass with no information on its radius. 

\begin{figure*}
    \centering
    \includegraphics[width=0.99\textwidth]{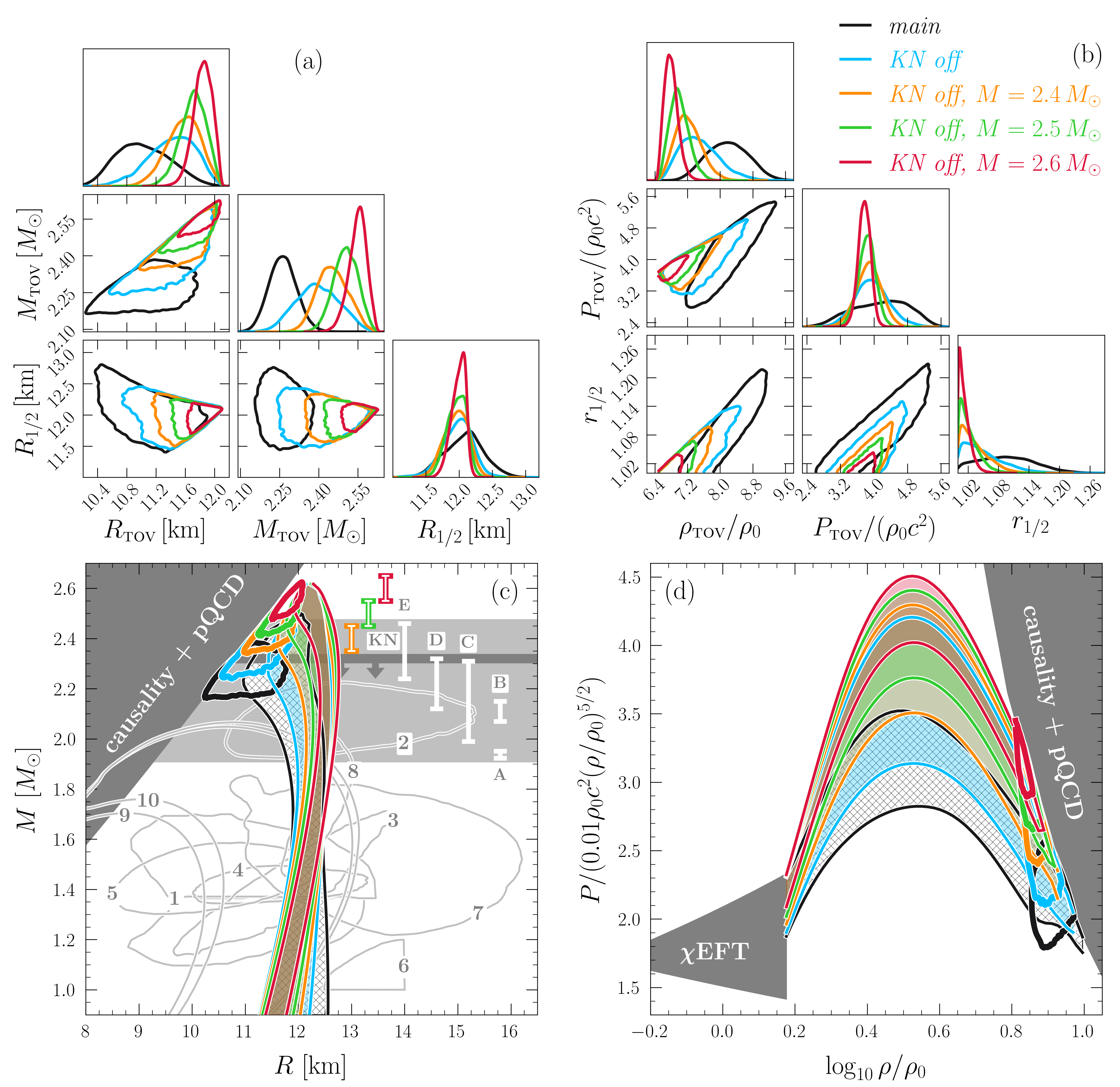}
    \caption{The same as Fig.~\ref{fig:PaperI-vs-basic} but for a comparison of the \basic\ scenario (black lines), the \textit{KN off} scenario with the upper limit from the EM counterpart of GW~170817~\cite{RezzollaMostWeihApJL2018} switched off (skyblue curves), and effects of a hypothetical observation of a NS with $2.4\pm 0.05\,\Msun$ (orange), $2.5\pm 0.05\,\Msun$ (green), or $2.6\pm 0.05\,\Msun$ (red) added to the \textit{KN off} data set. } 
    \label{fig:Mtest_24-25-26}
\end{figure*}

A mass measurement that is consistent with the \basic\ scenario can improve correspondent constraints on the EoS if only its uncertainties are several times better than the uncertainty of $\Mmax$ within this scenario. For instance, it can be an NS with $2.3\pm 0.01\,\Msun$ (hereafter in this section we use 68\% credibility levels for masses). But note that such masses are achievable in spider binaries only \cite[e.g.][]{MisraLinaresYeAA2025_Spiders}, mass measurements of which currently suffer from larger systematics \cite[e.g.][]{SullivanRomaniApJ2024_2215}. Instead, we consider a potential observation of an extremely massive NS with a reasonable accuracy $\pm 0.05\,\Msun$ (cf. PSR~J1810--1740 marked B in figures) that is in tension with the \basic\ scenario by exceeding the upper limit KN. The latter is based on an interpretation of the afterglow of GW170817, rather than on a direct mass measurement. We examine three options: $M = 2.4\pm 0.05\,\Msun$, 
$2.5\pm 0.05\,\Msun$, 
(both are at least marginally consistent with the KN limit)
and $2.6\pm 0.05\,\Msun$ (which is inconsistent). They are shown in Fig.~\ref{fig:Mtest_24-25-26}c with double-lined error bars. 
We add these hypothetical NSs to the \textit{KN off} scenario, which is the \basic\ data set with the KN limit excluded. Corresponding inferences for the EoS are shown in Fig.~\ref{fig:Mtest_24-25-26}.
 
Discarding the KN upper limit, we obtain essentially different inferences on the maximal mass NS parameters, and the ratio $\ratio$ is pressed against its assumed minimal value 1. This makes the $M-R$ credibility band somewhat narrower at low masses in the \textit{KN off} scenario than in the \basic\ case. 
However, eliminating the explicit upper limit on $\Mmax$ does not result in an unlimited broadening of the high-mass tail of $\Mmax$ posterior. Similarly to the \ron\ and \mimimi\ scenarios in Fig.~\ref{fig:Roff-Moff_main}, combination of causality, the constraint $\ratio > 1$, and the chiral corridor effectively constrains $\Mmax \lesssim 2.6\,\Msun$.

The same upper bound on $\Mmax$ holds for the data sets with an extremely massive NS added. An observation of this star just pushes the lower limit on $\Mmax$ upwards, prohibiting a part of the EoS parameter space allowed in the \textit{KN off} scenario. The remaining part of the parameter space is the smaller the higher the mass of the hypothetical NS is. As a consequence, the credibility bands for the $M-R$ and $P-\rho$ curves shrink along the whole range of masses and densities up to the TOV limit. For $2.6\,\Msun$ NS, the EoS inference is about 3 times narrower than for that of the \textit{KN off} scenario. So, depending on the mass, observing an extremely massive NS may meaningfully (but again, not dramatically) improve our knowledge of the EoS. However, within our approach, radii of NSs with $M < \Mmax/2$ decrease with increasing $\Mmax$, and some tension with the low-mass NS PSR~J1231--1411 (source 6 in Fig.~\ref{fig:Mtest_24-25-26}) emerges. Nevertheless, it is remarkable that a possible existence of an NS with $M \sim 2.6\,\Msun$ is still consistent with both current theoretical and most of the observational constraints on the NS EoS (except the radius of PSR~J1231--1411 and the KN upper limit on mass).

\section{Discussion \&\ Conclusions}
\label{sec:concl}

In this work, we examine the implications of current and future observational and theoretical constraints on the NS EoS, using a semi-analytic approach to IOVM. This approach is based on the 3-parametric representation of the EoS of NS cores, proposed in \paperi~\cite{OfPirShtPRD2024}.
Among the parameters determining the EoS, the TOV limit $\Mmax$ is the most crucial one. While the importance of $\Mmax$ is widely recognized, our 3-parametric approach appears to be an effective way to quantify this fact.
In accordance with the results of~\cite{Koehn+PRX2025}, our investigations show that the principal constraints on the EoS follow predominantly from the fundamental theoretical limits ($\chi$EFT at low densities, causality+pQCD for the TOV limit), supplemented by the minimal astrophysical information that $\Mmax$ should be larger than $\sim 2\,\Msun$ (see Fig.~\ref{fig:Roff-Moff_main} and the subsequent discussion). 
All other measurements of NS masses and radii---at least at the current accuracy level---improve the EoS inference quantitatively but not qualitatively. While most of the radius measurements constrain the EoS only at low densities, observations that limit the maximal NS mass have an impact at almost all densities. 

Based on our study of the possible impact of future NS observations, our conclusions underscore the fundamental difficulty of significantly improving EoS constraints through any single measurement. We find that neither an extremely precise single-radius measurement nor even the discovery of a very massive star (e.g., $M \sim 2.6\pm 0.05\,\Msun$; see Fig.~\ref{fig:Mtest_24-25-26}) would change dramatically our current EoS constraints. 

Nevertheless, our studies show that the most effective observation that will improve our knowledge of the EoS is one that will enable us to decrease the allowed region of the $(\Mmax,\Rmax)$ parameter subspace. For example, a discovery of an extremely massive NS with $M \gtrsim 2.5\,\Msun$ would have such an effect. However, even if the EoS allows such a massive NS, the possibility of such a detection is questionable due to potential evolutionary limitations on the formation of such a massive NS~\cite{MisraLinaresYeAA2025_Spiders}. A precise ($\pm 0.01\,\Msun$ or less) observation of a NS with $M > 2.2\Msun$ could also have such an impact. But the required precision seems to be unrealistic. 

A particularly promising possibility is to place an upper limit on $\Mmax$ using some indirect observations~\cite{RezzollaMostWeihApJL2018,NathanailMostRezzollaApJL2021,PernaGottlieb+PRD2025,ChenGottlieb2025_Mmax}. Given the significance of these constraints, it is of utmost importance to sharpen them and, at the same time,  to rule out alternative interpretations of the observations that could evade them.

Observations of individual NS radii, if consistent with the current EoS inference may predominantly improve the constraints on the EoS at low densities less than or approximately $2-3\rho_0$ but have no significant effect at higher densities. However,  considerably low or large radii may have important effect as they may imply new physics, as discussed below. 

In this work, we have discussed the implications of mass and radius measurements. Those are not the only NS characteristics that may be measured to constrain the EoS. In particular, tidal deformabilities can be extracted from binary NS merger observations \cite{GW170817,Chatziioannou+ArXiv2024} and there are proposals for measuring the moment of inertia \cite{LattimerSchutzApJ2005} and eigenfrequencies of NS oscillation modes~\cite{Iacovelli+PRD2023_NucPhysET}. These quantities depend on the NS EoS and can be used to infer it. However, using the universal relations between these quantities and NS masses and radii \cite{LattimerPrakashApJ2001, YagiYounesSci2013,AnderssonKokkotasMNRAS1998}, one can transfer our results about $M$ and $R$ measurements to any other type of future single star observation. 

While a single observation of an NS seems to be ineffective for further constraining the EoS, multiple observations can have a combined statistical effect. This is in line with prospects for EoS determination using next-generation GW detectors such as the Einstein Telescope~\cite{Iacovelli+PRD2023_NucPhysET}, which indicate that multiple merger detections (dozens or hundreds) are needed to pin down the EoS. The efficient methodology proposed here will be particularly useful in analyzing the implications of numerous such observations. 

To conclude, we note that we must keep in mind the possibility that some future observation might be strongly inconsistent with the current constraints on the NS EoS. This will break the picture described here and indicate that some new physics takes place inside NS cores. Naturally, the methods described here will become inapplicable and will have to be revised. For instance, detecting an NS with a radius less than $10\,$km will likely mean that there are two generations of compact stars with two distinct branches of the $M-R$ curve. This is possible if a strong first-order phase transition occurs in the NS core. Another possible groundbreaking observation would be a slowly rotating NS with $M > 2.7\,\Msun$, which will be in tension with either $\chi$EFT results for the low-density EoS or the fundamental causality condition. These exciting prospects, that we didn't consider here,  underscore the continued importance of diverse neutron-star observations for constraining the EoS.

\acknowledgements
We thank S.~Reddy, J.~Krolik, S.~Ginzburg, and S.~A.~Balashev for fruitful discussions and useful advice.
Work of DO and TP is supported by and advanced ERC grant MultiJets and by the Simons Collaboration on Extreme Electrodynamics of Compact Sources (SCEECS). 
Work of PS was supported by the Russian Science Foundation (grant 24-12-00320).

%

\appendix

\begin{figure*}
    \centering
    \includegraphics[width=\textwidth]{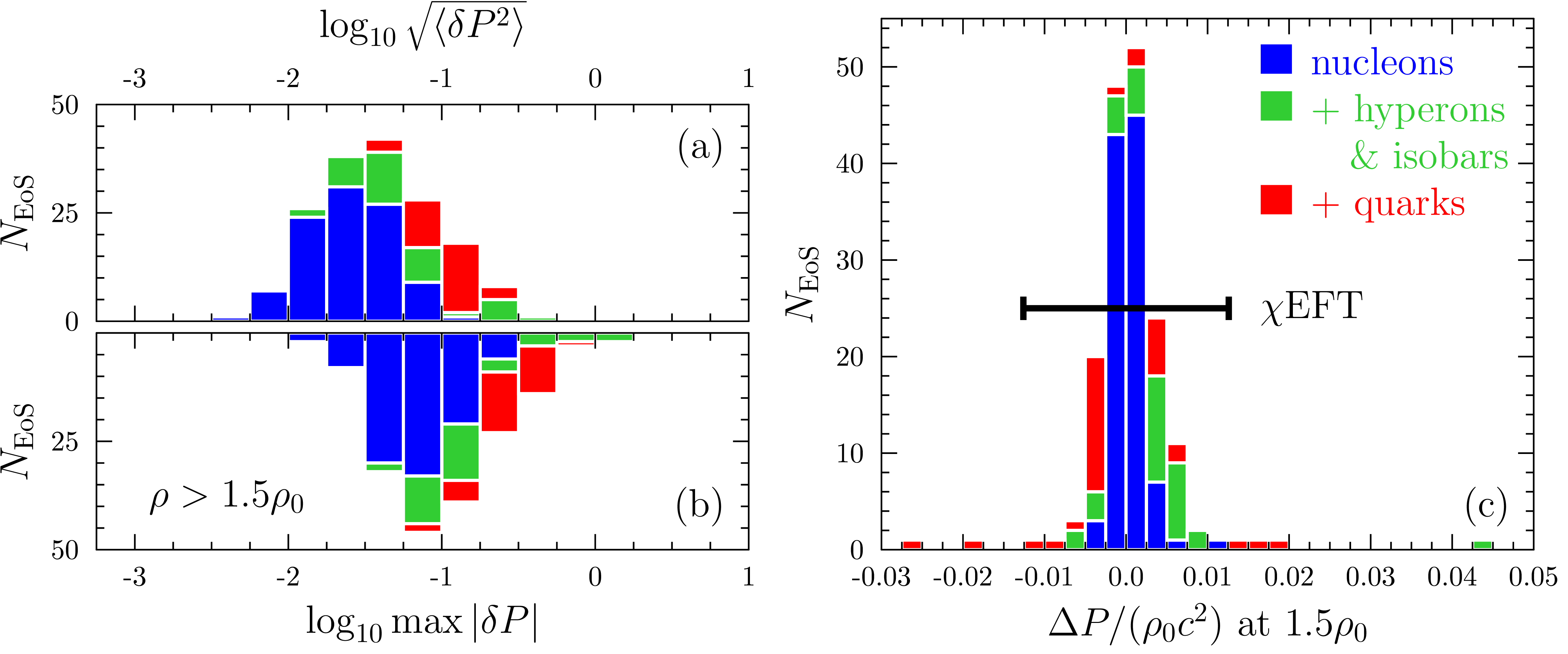}
    \caption{Distributions of relative root mean square error (a) and maximum relative errors (b) for Eq.~\eqref{eq:Prhofit} over the whole EoS collection. In panel (c), the distribution of absolute errors at $\rho=1.5\rho_0$ is compared to the width of the $\chi$EFT corridor (black bar) at the same density.}
    \label{fig:errorsPrho}
\end{figure*}

\section{Three-parametric representation of NS EoS}
\label{sec:app:approxes}

Here we list the set of approximations that give the three-parametric representation of the NS EoS and allow one to perform the explicit IOVM from a $M-R$ curve in the range $M = 1\,\Msun...\Mmax$ to a $P-\rho$ dependence in the range $\rho = 1.5\rho_0...\rhomax$. Eqs.~\eqref{eq:maxCorrs} and~\eqref{eq:MRfit} are the same as in \paperi, Eqs.~\eqref{eq:Prhofit} have the same functional form but the coefficients are optimized for the relevant density range.

Auxiliary functions---correlations of the properties of maximum mass NS:
\begin{subequations}\label{eq:maxCorrs}
\begin{align}
    \Mmax &= \sqrt{\frac{\Pmax^3}{G^3 \rhomax^4}} \frac{1}{f_M(\Pmax,\rhomax)},
    \label{eq:maxCorrs:M}\\
    \Rmax &= \sqrt{\frac{\Pmax}{G \rhomax^2}} \frac{1}{f_R(\Pmax,\rhomax)}, 
    \label{eq:maxCorrs:R}\\
    \csmax &= \sqrt{\frac{\Pmax}{\rhomax}}\frac{f_c(\Pmax,\rhomax)}{f_R(\Pmax,\rhomax)} = c\ \zeta(\Pmax,\rhomax) \ . \label{eq:maxCorrs:c}
\end{align}
Here $G$ is the gravitational constant, and $c$ is the speed of light. The dimensionless functions $f_M$, $f_R$, and $f_c$ have the same form
\begin{equation}\label{eq:maxCorr-f}
    f = C \left(\frac{\Pmax}{\rho_0c^2}\right)^{a} \left(\frac{\rhomax}{\rho_0}\right)^{b} + D
\end{equation}
with 
$\{a,b,C,D\} = \{1.41,-1.39,5.86,0.27\}$ for $f_M$, 
$\{0.744,-0.839,2.45,0.357\}$ for $f_R$, and 
$\{0.985,-0.990,3.02,0.529\}$ for $f_c$. 
\end{subequations}

Approximation for the $M-R$ relation in the range $1\,\Msun...\Mmax$:
\begin{multline}
    \label{eq:MRfit}
    R = \Rmax\left\{ 1 + \left[ 2\left(\sqrt{2}-1\right)\ratio - a \right]\sqrt{1-\frac{M}{\Mmax}} \right.\\
    \left. + \left[ 2\left(\sqrt{2}-1\right)\ratio - 2 + a\sqrt{2} \right]\left( 1-\frac{M}{\Mmax} \right) \right\},
\end{multline}
where $a=0.492$, and $\Mmax$ and $\Rmax$ depend on $\Pmax$ and $\rhomax$ via Eqs.~\eqref{eq:maxCorrs:M} and~\eqref{eq:maxCorrs:R}.

Approximation for the $P-\rho$ relation in the range $1.5\rho_0...\rhomax$:
\begin{subequations}\label{eq:Prhofit}
\begin{align}
    P &= \Pmax g_1 e^{g_2}, \label{eq:Prhofit:tot}\\
    g_1 &= \frac{\left( \rho/\rhomax \right)^{\rhomax\csmax^2/\Pmax-a_0}}{1 + a_0 ( 1 -
     \rho/\rhomax)  
     + (b_0 + b_1 \csmax/c)\left( 1 - \rho/\rhomax \right)^p } \label{eq:Prhofit:high}\\
     g_2 &= \left( u_0 + u_1 \ratio + u_2\ratio^2 \right) \left( e^{-v\rho/\rho_0} - e^{-v\rhomax/\rho_0} \right)\label{eq:Prhofit:low}
\end{align}
\end{subequations}
where 
$a_0 = -0.392$, 
$b_0 = -0.869$,
$b_1 = 1.69$, 
$p = 2.98$,
$u_0 = -60.9$,
$u_1 = 73.9$,
$u_2 = -21.4$,
$v = 1.50$.

Accuracy of the approximations \eqref{eq:maxCorrs} and \eqref{eq:MRfit} can be found in \paperi. The approximation~\eqref{eq:Prhofit} has overall accuracy slightly better than the approximation for the range $\rho>1\rho_0$ in \paperi\ (cf. Figs.~\ref{fig:errorsPrho}a,b and figures~4e,f there). An important improvement is that errors of the new approximation at the end of the chiral corridor, $\rho=1.5\rho_0$, are smaller than the width of the corridor (Fig.~\ref{fig:errorsPrho}c).

\section{Including NANOGrav 15 yr data}
\label{sec:app:NNG15}
\begin{figure*}
    \centering
    \includegraphics[width=0.99\textwidth]{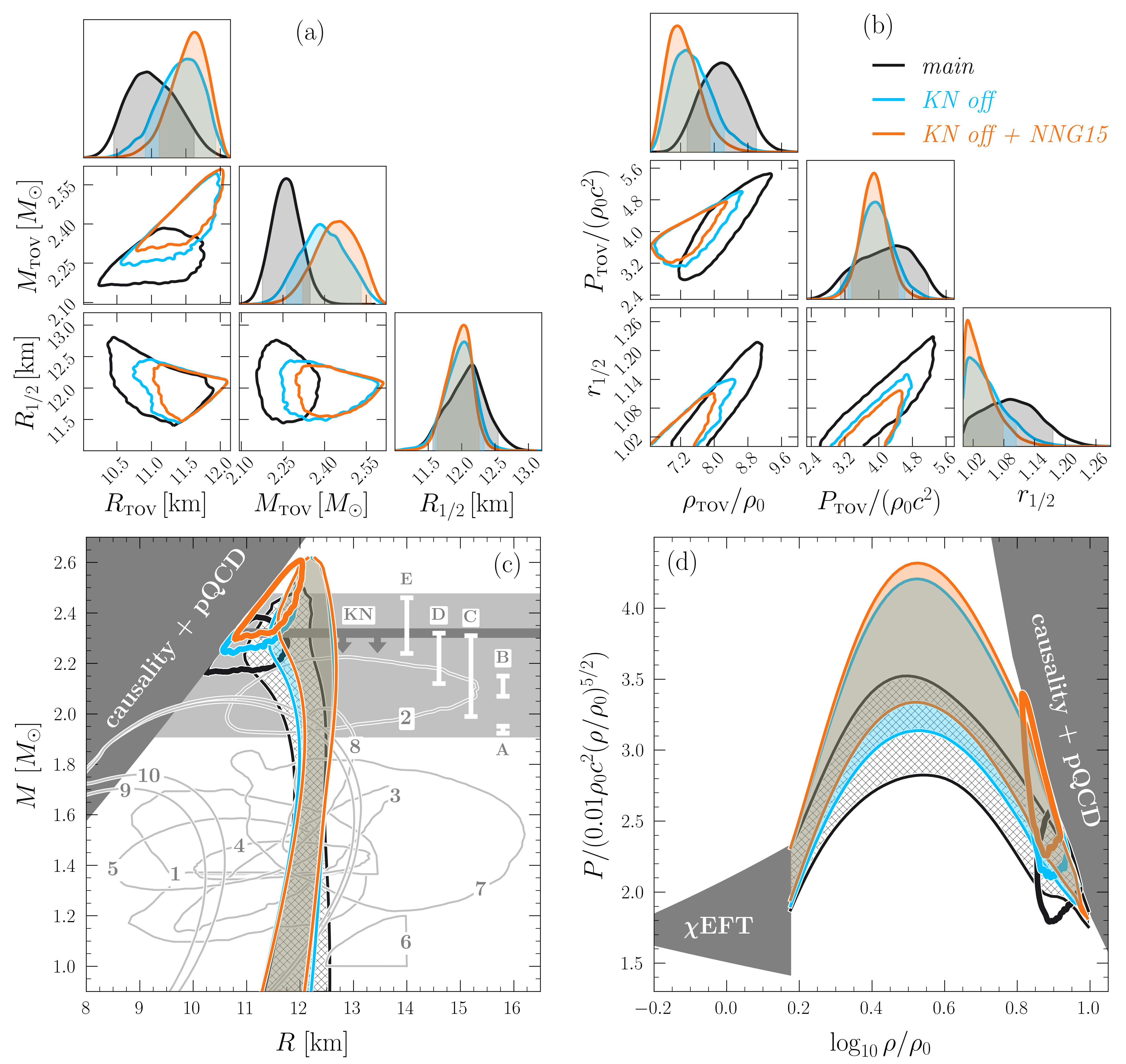}
    \caption{The same as Fig.~\ref{fig:PaperI-vs-basic} but for comparison of the \basic\ scenario (black lines), the \textit{KN off} scenario with the upper limit from the EM counterpart of GW~170817~\cite{RezzollaMostWeihApJL2018} switched off (skyblue curves), and the \textit{KN off + NNG15} scenario without the KN upper limit but with addition of the constraints from Shapiro-delay-derived mass measurements by NANOGrav 15 year data release~\cite{NANOGrav15years} (orange curves).} 
    \label{fig:Mtest_NNG15}
\end{figure*}

The 15 years data release by NANOGRav \cite{NANOGrav15years} includes Shapiro-delay-derived mass measurements for 17 millisecond pulsars. Two of those, PSR~J1614--2230 and PSR~J0740+6620, are already included in our analysis (Tables~\ref{tab:updates}, \ref{tab:Moff-Roff}). Among the rest, we take nine pulsars that have a significant fraction  of their mass posteriors residing at $M>2\,\Msun$, namely 
PSRs~J0613--0200, 
J1600--3053, 
J1630+3734, 
J1640+2224, 
J1811--2405, 
J1853+1303, 
J1946+3417, 
J2017+0603, 
and J2302+4442. 
Of course, some of these objects have unreasonably wide mass posteriors. For instance, the current estimate of the PSR~J1630+3734 mass is $6.3^{+4.2}_{-2.6}\,\Msun$ (1$\sigma$ credibility level). Such masses will likely be revised to lower values as the exposure time increases in future observations. Here, we consider the current measurements as a preliminary result. As $\Mmax$ can  barely exceed $3\,\Msun$ from theoretical grounds \cite{RhoadesRuffini1974}, we truncate the NANOGrav posteriors at $M=3\,\Msun$. 

As in Section~\ref{sec:future:m}, we study the impact of the NANOGrav data by adding it to the \textit{KN off} scenario, not the \basic\ one, since it is inconsistent with the KN upper limit. We demonstrate the impact of NANOGrav in Fig.~\ref{fig:Mtest_NNG15}. It is structured similarly to Figs.~\ref{fig:PaperI-vs-basic}, \ref{fig:Roff-Moff_main}, and~\ref{fig:Mtest_24-25-26}, containing three EoS fits to the data: the \basic\ scenario, the \textit{KN off} scenario that is \basic\ without the upper limit from the EM counterpart to GW170817, and the \textit{KN off + NNG15} scenario that is the previous one plus nine NS mass estimates from the NANOGrav 15 years data release~\cite{NANOGrav15years}. Due to rather large uncertainties, the collective mass distribution of the NANOGrav pulsars has a broad low-mass tail. As a result, EoS inferences for the \textit{KN off} and \textit{KN off + NNG15} scenarios show no statistically significant difference. 


\begin{thebibliography}{62}%
\makeatletter
\providecommand \@ifxundefined [1]{%
 \@ifx{#1\undefined}
}%
\providecommand \@ifnum [1]{%
 \ifnum #1\expandafter \@firstoftwo
 \else \expandafter \@secondoftwo
 \fi
}%
\providecommand \@ifx [1]{%
 \ifx #1\expandafter \@firstoftwo
 \else \expandafter \@secondoftwo
 \fi
}%
\providecommand \natexlab [1]{#1}%
\providecommand \enquote  [1]{``#1''}%
\providecommand \bibnamefont  [1]{#1}%
\providecommand \bibfnamefont [1]{#1}%
\providecommand \citenamefont [1]{#1}%
\providecommand \href@noop [0]{\@secondoftwo}%
\providecommand \href [0]{\begingroup \@sanitize@url \@href}%
\providecommand \@href[1]{\@@startlink{#1}\@@href}%
\providecommand \@@href[1]{\endgroup#1\@@endlink}%
\providecommand \@sanitize@url [0]{\catcode `\\12\catcode `\$12\catcode `\&12\catcode `\#12\catcode `\^12\catcode `\_12\catcode `\%12\relax}%
\providecommand \@@startlink[1]{}%
\providecommand \@@endlink[0]{}%
\providecommand \url  [0]{\begingroup\@sanitize@url \@url }%
\providecommand \@url [1]{\endgroup\@href {#1}{\urlprefix }}%
\providecommand \urlprefix  [0]{URL }%
\providecommand \Eprint [0]{\href }%
\providecommand \doibase [0]{https://doi.org/}%
\providecommand \selectlanguage [0]{\@gobble}%
\providecommand \bibinfo  [0]{\@secondoftwo}%
\providecommand \bibfield  [0]{\@secondoftwo}%
\providecommand \translation [1]{[#1]}%
\providecommand \BibitemOpen [0]{}%
\providecommand \bibitemStop [0]{}%
\providecommand \bibitemNoStop [0]{.\EOS\space}%
\providecommand \EOS [0]{\spacefactor3000\relax}%
\providecommand \BibitemShut  [1]{\csname bibitem#1\endcsname}%
\let\auto@bib@innerbib\@empty
\bibitem [{\citenamefont {{Lattimer}}(2021)}]{LattimerAnnRevNucPhys2021}%
  \BibitemOpen
  \bibfield  {author} {\bibinfo {author} {\bibfnamefont {J.~M.}\ \bibnamefont {{Lattimer}}},\ }\bibfield  {title} {\bibinfo {title} {{Neutron Stars and the Nuclear Matter Equation of State}},\ }\href {https://doi.org/10.1146/annurev-nucl-102419-124827} {\bibfield  {journal} {\bibinfo  {journal} {Annual Review of Nuclear and Particle Science}\ }\textbf {\bibinfo {volume} {71}},\ \bibinfo {pages} {433} (\bibinfo {year} {2021})}\BibitemShut {NoStop}%
\bibitem [{\citenamefont {{Chatziioannou}}\ \emph {et~al.}(2025)\citenamefont {{Chatziioannou}}, \citenamefont {{Cromartie}}, \citenamefont {{Gandolfi}}, \citenamefont {{Tews}}, \citenamefont {{Radice}}, \citenamefont {{Steiner}},\ and\ \citenamefont {{Watts}}}]{Chatziioannou+ArXiv2024}%
  \BibitemOpen
  \bibfield  {author} {\bibinfo {author} {\bibfnamefont {K.}~\bibnamefont {{Chatziioannou}}}, \bibinfo {author} {\bibfnamefont {H.~T.}\ \bibnamefont {{Cromartie}}}, \bibinfo {author} {\bibfnamefont {S.}~\bibnamefont {{Gandolfi}}}, \bibinfo {author} {\bibfnamefont {I.}~\bibnamefont {{Tews}}}, \bibinfo {author} {\bibfnamefont {D.}~\bibnamefont {{Radice}}}, \bibinfo {author} {\bibfnamefont {A.~W.}\ \bibnamefont {{Steiner}}},\ and\ \bibinfo {author} {\bibfnamefont {A.~L.}\ \bibnamefont {{Watts}}},\ }\bibfield  {title} {\bibinfo {title} {{Neutron stars and the dense matter equation of state}},\ }\href {https://doi.org/10.1103/ymsq-cfcw} {\bibfield  {journal} {\bibinfo  {journal} {Reviews of Modern Physics}\ }\textbf {\bibinfo {volume} {97}},\ \bibinfo {eid} {045007} (\bibinfo {year} {2025})},\ \Eprint {https://arxiv.org/abs/2407.11153} {arXiv:2407.11153 [nucl-th]} \BibitemShut {NoStop}%
\bibitem [{\citenamefont {{Ofengeim}}\ \emph {et~al.}(2026)\citenamefont {{Ofengeim}}, \citenamefont {{Kolomeitsev}},\ and\ \citenamefont {{Shternin}}}]{TheChapter}%
  \BibitemOpen
  \bibfield  {author} {\bibinfo {author} {\bibfnamefont {D.~D.}\ \bibnamefont {{Ofengeim}}}, \bibinfo {author} {\bibfnamefont {E.~E.}\ \bibnamefont {{Kolomeitsev}}},\ and\ \bibinfo {author} {\bibfnamefont {P.~S.}\ \bibnamefont {{Shternin}}},\ }\bibfield  {title} {\bibinfo {title} {{Neutron star equation of state}},\ }in\ \href {https://doi.org/10.1016/B978-0-443-21439-4.00137-1} {\emph {\bibinfo {booktitle} {Encyclopedia of Astrophysics, Volume 3}}},\ Vol.~\bibinfo {volume} {3}\ (\bibinfo {year} {2026})\ pp.\ \bibinfo {pages} {160--204}\BibitemShut {NoStop}%
\bibitem [{\citenamefont {{Tolman}}(1939)}]{Tolman1939}%
  \BibitemOpen
  \bibfield  {author} {\bibinfo {author} {\bibfnamefont {R.~C.}\ \bibnamefont {{Tolman}}},\ }\bibfield  {title} {\bibinfo {title} {{Static Solutions of Einstein's Field Equations for Spheres of Fluid}},\ }\href {https://doi.org/10.1103/PhysRev.55.364} {\bibfield  {journal} {\bibinfo  {journal} {Physical Review}\ }\textbf {\bibinfo {volume} {55}},\ \bibinfo {pages} {364} (\bibinfo {year} {1939})}\BibitemShut {NoStop}%
\bibitem [{\citenamefont {{Oppenheimer}}\ and\ \citenamefont {{Volkoff}}(1939)}]{OppVol1939}%
  \BibitemOpen
  \bibfield  {author} {\bibinfo {author} {\bibfnamefont {J.~R.}\ \bibnamefont {{Oppenheimer}}}\ and\ \bibinfo {author} {\bibfnamefont {G.~M.}\ \bibnamefont {{Volkoff}}},\ }\bibfield  {title} {\bibinfo {title} {{On Massive Neutron Cores}},\ }\href {https://doi.org/10.1103/PhysRev.55.374} {\bibfield  {journal} {\bibinfo  {journal} {Physical Review}\ }\textbf {\bibinfo {volume} {55}},\ \bibinfo {pages} {374} (\bibinfo {year} {1939})}\BibitemShut {NoStop}%
\bibitem [{\citenamefont {{Lindblom}}(1992)}]{Lind1992}%
  \BibitemOpen
  \bibfield  {author} {\bibinfo {author} {\bibfnamefont {L.}~\bibnamefont {{Lindblom}}},\ }\bibfield  {title} {\bibinfo {title} {{Determining the Nuclear Equation of State from Neutron-Star Masses and Radii}},\ }\href {https://doi.org/10.1086/171882} {\bibfield  {journal} {\bibinfo  {journal} {\apj}\ }\textbf {\bibinfo {volume} {398}},\ \bibinfo {pages} {569} (\bibinfo {year} {1992})}\BibitemShut {NoStop}%
\bibitem [{\citenamefont {{Lattimer}}\ and\ \citenamefont {{Prakash}}(2001)}]{LattimerPrakashApJ2001}%
  \BibitemOpen
  \bibfield  {author} {\bibinfo {author} {\bibfnamefont {J.~M.}\ \bibnamefont {{Lattimer}}}\ and\ \bibinfo {author} {\bibfnamefont {M.}~\bibnamefont {{Prakash}}},\ }\bibfield  {title} {\bibinfo {title} {{Neutron Star Structure and the Equation of State}},\ }\href {https://doi.org/10.1086/319702} {\bibfield  {journal} {\bibinfo  {journal} {\apj}\ }\textbf {\bibinfo {volume} {550}},\ \bibinfo {pages} {426} (\bibinfo {year} {2001})},\ \Eprint {https://arxiv.org/abs/astro-ph/0002232} {arXiv:astro-ph/0002232 [astro-ph]} \BibitemShut {NoStop}%
\bibitem [{\citenamefont {{Yagi}}\ and\ \citenamefont {{Yunes}}(2013{\natexlab{a}})}]{YagiYounesPRD2013}%
  \BibitemOpen
  \bibfield  {author} {\bibinfo {author} {\bibfnamefont {K.}~\bibnamefont {{Yagi}}}\ and\ \bibinfo {author} {\bibfnamefont {N.}~\bibnamefont {{Yunes}}},\ }\bibfield  {title} {\bibinfo {title} {{I-Love-Q relations in neutron stars and their applications to astrophysics, gravitational waves, and fundamental physics}},\ }\href {https://doi.org/10.1103/PhysRevD.88.023009} {\bibfield  {journal} {\bibinfo  {journal} {\prd}\ }\textbf {\bibinfo {volume} {88}},\ \bibinfo {eid} {023009} (\bibinfo {year} {2013}{\natexlab{a}})},\ \Eprint {https://arxiv.org/abs/1303.1528} {arXiv:1303.1528 [gr-qc]} \BibitemShut {NoStop}%
\bibitem [{\citenamefont {{Yagi}}\ and\ \citenamefont {{Yunes}}(2016)}]{YagiYunesCQG2016}%
  \BibitemOpen
  \bibfield  {author} {\bibinfo {author} {\bibfnamefont {K.}~\bibnamefont {{Yagi}}}\ and\ \bibinfo {author} {\bibfnamefont {N.}~\bibnamefont {{Yunes}}},\ }\bibfield  {title} {\bibinfo {title} {{Binary Love relations}},\ }\href {https://doi.org/10.1088/0264-9381/33/13/13LT01} {\bibfield  {journal} {\bibinfo  {journal} {Classical and Quantum Gravity}\ }\textbf {\bibinfo {volume} {33}},\ \bibinfo {eid} {13LT01} (\bibinfo {year} {2016})},\ \Eprint {https://arxiv.org/abs/1512.02639} {arXiv:1512.02639 [gr-qc]} \BibitemShut {NoStop}%
\bibitem [{\citenamefont {{Gerlach}}(1968)}]{GerlachPhysRev1968}%
  \BibitemOpen
  \bibfield  {author} {\bibinfo {author} {\bibfnamefont {U.~H.}\ \bibnamefont {{Gerlach}}},\ }\bibfield  {title} {\bibinfo {title} {{Equation of State at Supranuclear Densities and the Existence of a Third Family of Superdense Stars}},\ }\href {https://doi.org/10.1103/PhysRev.172.1325} {\bibfield  {journal} {\bibinfo  {journal} {Physical Review}\ }\textbf {\bibinfo {volume} {172}},\ \bibinfo {pages} {1325} (\bibinfo {year} {1968})}\BibitemShut {NoStop}%
\bibitem [{\citenamefont {{Foreman-Mackey}}\ \emph {et~al.}(2013)\citenamefont {{Foreman-Mackey}}, \citenamefont {{Hogg}}, \citenamefont {{Lang}},\ and\ \citenamefont {{Goodman}}}]{Foreman-Mackey+PASA2013_emcee}%
  \BibitemOpen
  \bibfield  {author} {\bibinfo {author} {\bibfnamefont {D.}~\bibnamefont {{Foreman-Mackey}}}, \bibinfo {author} {\bibfnamefont {D.~W.}\ \bibnamefont {{Hogg}}}, \bibinfo {author} {\bibfnamefont {D.}~\bibnamefont {{Lang}}},\ and\ \bibinfo {author} {\bibfnamefont {J.}~\bibnamefont {{Goodman}}},\ }\bibfield  {title} {\bibinfo {title} {{emcee: The MCMC Hammer}},\ }\href {https://doi.org/10.1086/670067} {\bibfield  {journal} {\bibinfo  {journal} {Publications of the Astronomical Society of the Pacific}\ }\textbf {\bibinfo {volume} {125}},\ \bibinfo {pages} {306} (\bibinfo {year} {2013})}\BibitemShut {NoStop}%
\bibitem [{\citenamefont {{Feroz}}\ \emph {et~al.}(2009)\citenamefont {{Feroz}}, \citenamefont {{Hobson}},\ and\ \citenamefont {{Bridges}}}]{Feroz+MNRAS_MULTINEST}%
  \BibitemOpen
  \bibfield  {author} {\bibinfo {author} {\bibfnamefont {F.}~\bibnamefont {{Feroz}}}, \bibinfo {author} {\bibfnamefont {M.~P.}\ \bibnamefont {{Hobson}}},\ and\ \bibinfo {author} {\bibfnamefont {M.}~\bibnamefont {{Bridges}}},\ }\bibfield  {title} {\bibinfo {title} {{MULTINEST: an efficient and robust Bayesian inference tool for cosmology and particle physics}},\ }\href {https://doi.org/10.1111/j.1365-2966.2009.14548.x} {\bibfield  {journal} {\bibinfo  {journal} {\mnras}\ }\textbf {\bibinfo {volume} {398}},\ \bibinfo {pages} {1601} (\bibinfo {year} {2009})},\ \Eprint {https://arxiv.org/abs/0809.3437} {arXiv:0809.3437 [astro-ph]} \BibitemShut {NoStop}%
\bibitem [{\citenamefont {{N{\"a}ttil{\"a}}}\ \emph {et~al.}(2016)\citenamefont {{N{\"a}ttil{\"a}}}, \citenamefont {{Steiner}}, \citenamefont {{Kajava}}, \citenamefont {{Suleimanov}},\ and\ \citenamefont {{Poutanen}}}]{Nattila+AA2016}%
  \BibitemOpen
  \bibfield  {author} {\bibinfo {author} {\bibfnamefont {J.}~\bibnamefont {{N{\"a}ttil{\"a}}}}, \bibinfo {author} {\bibfnamefont {A.~W.}\ \bibnamefont {{Steiner}}}, \bibinfo {author} {\bibfnamefont {J.~J.~E.}\ \bibnamefont {{Kajava}}}, \bibinfo {author} {\bibfnamefont {V.~F.}\ \bibnamefont {{Suleimanov}}},\ and\ \bibinfo {author} {\bibfnamefont {J.}~\bibnamefont {{Poutanen}}},\ }\bibfield  {title} {\bibinfo {title} {{Equation of state constraints for the cold dense matter inside neutron stars using the cooling tail method}},\ }\href {https://doi.org/10.1051/0004-6361/201527416} {\bibfield  {journal} {\bibinfo  {journal} {\aap}\ }\textbf {\bibinfo {volume} {591}},\ \bibinfo {eid} {A25} (\bibinfo {year} {2016})},\ \Eprint {https://arxiv.org/abs/1509.06561} {arXiv:1509.06561 [astro-ph.HE]} \BibitemShut {NoStop}%
\bibitem [{\citenamefont {{Annala}}\ \emph {et~al.}(2023)\citenamefont {{Annala}}, \citenamefont {{Gorda}}, \citenamefont {{Hirvonen}}, \citenamefont {{Komoltsev}}, \citenamefont {{Kurkela}},\ and\ \citenamefont {et~al.}}]{Annala+Nat2023}%
  \BibitemOpen
  \bibfield  {author} {\bibinfo {author} {\bibfnamefont {E.}~\bibnamefont {{Annala}}}, \bibinfo {author} {\bibfnamefont {T.}~\bibnamefont {{Gorda}}}, \bibinfo {author} {\bibfnamefont {J.}~\bibnamefont {{Hirvonen}}}, \bibinfo {author} {\bibfnamefont {O.}~\bibnamefont {{Komoltsev}}}, \bibinfo {author} {\bibfnamefont {A.}~\bibnamefont {{Kurkela}}},\ and\ \bibinfo {author} {\bibnamefont {et~al.}},\ }\bibfield  {title} {\bibinfo {title} {{Strongly interacting matter exhibits deconfined behavior in massive neutron stars}},\ }\href {https://doi.org/https://doi.org/10.1038/s41467-023-44051-y} {\bibfield  {journal} {\bibinfo  {journal} {Nat. Commun.}\ }\textbf {\bibinfo {volume} {14}},\ \bibinfo {eid} {arXiv:2303.11356} (\bibinfo {year} {2023})}\BibitemShut {NoStop}%
\bibitem [{\citenamefont {{Brandes}}\ \emph {et~al.}(2023)\citenamefont {{Brandes}}, \citenamefont {{Weise}},\ and\ \citenamefont {{Kaiser}}}]{Brandes+PRD2023}%
  \BibitemOpen
  \bibfield  {author} {\bibinfo {author} {\bibfnamefont {L.}~\bibnamefont {{Brandes}}}, \bibinfo {author} {\bibfnamefont {W.}~\bibnamefont {{Weise}}},\ and\ \bibinfo {author} {\bibfnamefont {N.}~\bibnamefont {{Kaiser}}},\ }\bibfield  {title} {\bibinfo {title} {{Evidence against a strong first-order phase transition in neutron star cores: Impact of new data}},\ }\href {https://doi.org/10.1103/PhysRevD.108.094014} {\bibfield  {journal} {\bibinfo  {journal} {\prd}\ }\textbf {\bibinfo {volume} {108}},\ \bibinfo {eid} {094014} (\bibinfo {year} {2023})},\ \Eprint {https://arxiv.org/abs/2306.06218} {arXiv:2306.06218 [nucl-th]} \BibitemShut {NoStop}%
\bibitem [{\citenamefont {{Rutherford}}\ \emph {et~al.}(2024)\citenamefont {{Rutherford}}, \citenamefont {{Mendes}}, \citenamefont {{Svensson}}, \citenamefont {{Schwenk}}, \citenamefont {{Watts}},\ and\ \citenamefont {et~al.}}]{Rutherford2024}%
  \BibitemOpen
  \bibfield  {author} {\bibinfo {author} {\bibfnamefont {N.}~\bibnamefont {{Rutherford}}}, \bibinfo {author} {\bibfnamefont {M.}~\bibnamefont {{Mendes}}}, \bibinfo {author} {\bibfnamefont {I.}~\bibnamefont {{Svensson}}}, \bibinfo {author} {\bibfnamefont {A.}~\bibnamefont {{Schwenk}}}, \bibinfo {author} {\bibfnamefont {A.~L.}\ \bibnamefont {{Watts}}},\ and\ \bibinfo {author} {\bibnamefont {et~al.}},\ }\bibfield  {title} {\bibinfo {title} {{Constraining the Dense Matter Equation of State with New NICER Mass{\textendash}Radius Measurements and New Chiral Effective Field Theory Inputs}},\ }\href {https://doi.org/10.3847/2041-8213/ad5f02} {\bibfield  {journal} {\bibinfo  {journal} {\apjl}\ }\textbf {\bibinfo {volume} {971}},\ \bibinfo {eid} {L19} (\bibinfo {year} {2024})},\ \Eprint {https://arxiv.org/abs/2407.06790} {arXiv:2407.06790 [astro-ph.HE]} \BibitemShut {NoStop}%
\bibitem [{\citenamefont {{Brandes}}\ and\ \citenamefont {{Weise}}(2025)}]{BrandesWeisePRD2025}%
  \BibitemOpen
  \bibfield  {author} {\bibinfo {author} {\bibfnamefont {L.}~\bibnamefont {{Brandes}}}\ and\ \bibinfo {author} {\bibfnamefont {W.}~\bibnamefont {{Weise}}},\ }\bibfield  {title} {\bibinfo {title} {{Implications of latest NICER data for the neutron star equation of state}},\ }\href {https://doi.org/10.1103/PhysRevD.111.034005} {\bibfield  {journal} {\bibinfo  {journal} {\prd}\ }\textbf {\bibinfo {volume} {111}},\ \bibinfo {eid} {034005} (\bibinfo {year} {2025})},\ \Eprint {https://arxiv.org/abs/2412.05923} {arXiv:2412.05923 [nucl-th]} \BibitemShut {NoStop}%
\bibitem [{\citenamefont {{Koehn}}\ \emph {et~al.}(2025)\citenamefont {{Koehn}}, \citenamefont {{Rose}}, \citenamefont {{Pang}}, \citenamefont {{Somasundaram}}, \citenamefont {{Reed}},\ and\ \citenamefont {et~al.}}]{Koehn+PRX2025}%
  \BibitemOpen
  \bibfield  {author} {\bibinfo {author} {\bibfnamefont {H.}~\bibnamefont {{Koehn}}}, \bibinfo {author} {\bibfnamefont {H.}~\bibnamefont {{Rose}}}, \bibinfo {author} {\bibfnamefont {P.~T.~H.}\ \bibnamefont {{Pang}}}, \bibinfo {author} {\bibfnamefont {R.}~\bibnamefont {{Somasundaram}}}, \bibinfo {author} {\bibfnamefont {B.~T.}\ \bibnamefont {{Reed}}},\ and\ \bibinfo {author} {\bibnamefont {et~al.}},\ }\bibfield  {title} {\bibinfo {title} {{From Existing and New Nuclear and Astrophysical Constraints to Stringent Limits on the Equation of State of Neutron-Rich Dense Matter}},\ }\href {https://doi.org/10.1103/PhysRevX.15.021014} {\bibfield  {journal} {\bibinfo  {journal} {Physical Review X}\ }\textbf {\bibinfo {volume} {15}},\ \bibinfo {eid} {021014} (\bibinfo {year} {2025})},\ \Eprint {https://arxiv.org/abs/2402.04172} {arXiv:2402.04172 [astro-ph.HE]} \BibitemShut {NoStop}%
\bibitem [{\citenamefont {{Komoltsev}}(2024)}]{KomoltsevPRD2024_FOPT}%
  \BibitemOpen
  \bibfield  {author} {\bibinfo {author} {\bibfnamefont {O.}~\bibnamefont {{Komoltsev}}},\ }\bibfield  {title} {\bibinfo {title} {{First-order phase transitions in the cores of neutron stars}},\ }\href {https://doi.org/10.1103/PhysRevD.110.L071502} {\bibfield  {journal} {\bibinfo  {journal} {\prd}\ }\textbf {\bibinfo {volume} {110}},\ \bibinfo {eid} {L071502} (\bibinfo {year} {2024})},\ \Eprint {https://arxiv.org/abs/2404.05637} {arXiv:2404.05637 [nucl-th]} \BibitemShut {NoStop}%
\bibitem [{\citenamefont {{Tang}}\ \emph {et~al.}(2025)\citenamefont {{Tang}}, \citenamefont {{Huang}},\ and\ \citenamefont {{Fan}}}]{TangHuangFanPRD2025}%
  \BibitemOpen
  \bibfield  {author} {\bibinfo {author} {\bibfnamefont {S.-P.}\ \bibnamefont {{Tang}}}, \bibinfo {author} {\bibfnamefont {Y.-J.}\ \bibnamefont {{Huang}}},\ and\ \bibinfo {author} {\bibfnamefont {Y.-Z.}\ \bibnamefont {{Fan}}},\ }\bibfield  {title} {\bibinfo {title} {{Phase transition and nuclear symmetry energy from neutron star observations: Constraints in light of PSR J0614-3329}},\ }\href {https://doi.org/10.1103/bmsk-8n85} {\bibfield  {journal} {\bibinfo  {journal} {\prd}\ }\textbf {\bibinfo {volume} {112}},\ \bibinfo {eid} {083009} (\bibinfo {year} {2025})},\ \Eprint {https://arxiv.org/abs/2507.10025} {arXiv:2507.10025 [astro-ph.HE]} \BibitemShut {NoStop}%
\bibitem [{\citenamefont {{Ecker}}\ \emph {et~al.}(2026)\citenamefont {{Ecker}}, \citenamefont {{Jokela}},\ and\ \citenamefont {{J{\"a}rvinen}}}]{EckerJokelaJarvinenArXiv2025}%
  \BibitemOpen
  \bibfield  {author} {\bibinfo {author} {\bibfnamefont {C.}~\bibnamefont {{Ecker}}}, \bibinfo {author} {\bibfnamefont {N.}~\bibnamefont {{Jokela}}},\ and\ \bibinfo {author} {\bibfnamefont {M.}~\bibnamefont {{J{\"a}rvinen}}},\ }\bibfield  {title} {\bibinfo {title} {{Locating the QCD critical point with input from neutron-star observations}},\ }\href {https://doi.org/10.1103/x17s-sc9t} {\bibfield  {journal} {\bibinfo  {journal} {\prd}\ }\textbf {\bibinfo {volume} {113}},\ \bibinfo {eid} {L041302} (\bibinfo {year} {2026})},\ \Eprint {https://arxiv.org/abs/2506.10065} {arXiv:2506.10065 [astro-ph.HE]} \BibitemShut {NoStop}%
\bibitem [{\citenamefont {{Cuceu}}\ and\ \citenamefont {{Robles}}(2025)}]{CuceuRoblesPRD2025}%
  \BibitemOpen
  \bibfield  {author} {\bibinfo {author} {\bibfnamefont {I.}~\bibnamefont {{Cuceu}}}\ and\ \bibinfo {author} {\bibfnamefont {S.}~\bibnamefont {{Robles}}},\ }\bibfield  {title} {\bibinfo {title} {{Direct nonparametric multimessenger constraints on the equation of state of cold dense nuclear matter}},\ }\href {https://doi.org/10.1103/6d9c-c4kx} {\bibfield  {journal} {\bibinfo  {journal} {\prd}\ }\textbf {\bibinfo {volume} {111}},\ \bibinfo {eid} {123029} (\bibinfo {year} {2025})},\ \Eprint {https://arxiv.org/abs/2410.23407} {arXiv:2410.23407 [astro-ph.HE]} \BibitemShut {NoStop}%
\bibitem [{\citenamefont {{Biswas}}(2025)}]{BiswasArXiv2025}%
  \BibitemOpen
  \bibfield  {author} {\bibinfo {author} {\bibfnamefont {B.}~\bibnamefont {{Biswas}}},\ }\bibfield  {title} {\bibinfo {title} {{Equation of State Extrapolation Systematics: Parametric vs. Nonparametric Inference of Neutron Star Structure}},\ }\href {https://doi.org/10.48550/arXiv.2509.06145} {\bibfield  {journal} {\bibinfo  {journal} {arXiv e-prints}\ ,\ \bibinfo {eid} {arXiv:2509.06145}} (\bibinfo {year} {2025})},\ \Eprint {https://arxiv.org/abs/2509.06145} {arXiv:2509.06145 [astro-ph.HE]} \BibitemShut {NoStop}%
\bibitem [{\citenamefont {{Lindblom}}(2010)}]{Lindblom2010}%
  \BibitemOpen
  \bibfield  {author} {\bibinfo {author} {\bibfnamefont {L.}~\bibnamefont {{Lindblom}}},\ }\bibfield  {title} {\bibinfo {title} {{Spectral representations of neutron-star equations of state}},\ }\href {https://doi.org/10.1103/PhysRevD.82.103011} {\bibfield  {journal} {\bibinfo  {journal} {\prd}\ }\textbf {\bibinfo {volume} {82}},\ \bibinfo {eid} {103011} (\bibinfo {year} {2010})},\ \Eprint {https://arxiv.org/abs/1009.0738} {arXiv:1009.0738 [astro-ph.HE]} \BibitemShut {NoStop}%
\bibitem [{\citenamefont {{Brandes}}\ \emph {et~al.}(2024)\citenamefont {{Brandes}}, \citenamefont {{Modi}}, \citenamefont {{Ghosh}}, \citenamefont {{Farrell}}, \citenamefont {{Lindblom}},\ and\ \citenamefont {et~al.}}]{BrandesWeiseJCAP2024}%
  \BibitemOpen
  \bibfield  {author} {\bibinfo {author} {\bibfnamefont {L.}~\bibnamefont {{Brandes}}}, \bibinfo {author} {\bibfnamefont {C.}~\bibnamefont {{Modi}}}, \bibinfo {author} {\bibfnamefont {A.}~\bibnamefont {{Ghosh}}}, \bibinfo {author} {\bibfnamefont {D.}~\bibnamefont {{Farrell}}}, \bibinfo {author} {\bibfnamefont {L.}~\bibnamefont {{Lindblom}}},\ and\ \bibinfo {author} {\bibnamefont {et~al.}},\ }\bibfield  {title} {\bibinfo {title} {{Neural simulation-based inference of the neutron star equation of state directly from telescope spectra}},\ }\href {https://doi.org/10.1088/1475-7516/2024/09/009} {\bibfield  {journal} {\bibinfo  {journal} {\jcap}\ }\textbf {\bibinfo {volume} {2024}},\ \bibinfo {eid} {009} (\bibinfo {year} {2024})},\ \Eprint {https://arxiv.org/abs/2403.00287} {arXiv:2403.00287 [astro-ph.HE]} \BibitemShut {NoStop}%
\bibitem [{\citenamefont {{Soma}}\ \emph {et~al.}(2022)\citenamefont {{Soma}}, \citenamefont {{Wang}}, \citenamefont {{Shi}}, \citenamefont {{St{\"o}cker}}, \citenamefont {{Zhou}},\ and\ \citenamefont {et~al.}}]{Soma+JCAP2022}%
  \BibitemOpen
  \bibfield  {author} {\bibinfo {author} {\bibfnamefont {S.}~\bibnamefont {{Soma}}}, \bibinfo {author} {\bibfnamefont {L.}~\bibnamefont {{Wang}}}, \bibinfo {author} {\bibfnamefont {S.}~\bibnamefont {{Shi}}}, \bibinfo {author} {\bibfnamefont {H.}~\bibnamefont {{St{\"o}cker}}}, \bibinfo {author} {\bibfnamefont {K.}~\bibnamefont {{Zhou}}},\ and\ \bibinfo {author} {\bibnamefont {et~al.}},\ }\bibfield  {title} {\bibinfo {title} {{Neural network reconstruction of the dense matter equation of state from neutron star observables}},\ }\href {https://doi.org/10.1088/1475-7516/2022/08/071} {\bibfield  {journal} {\bibinfo  {journal} {\jcap}\ }\textbf {\bibinfo {volume} {2022}},\ \bibinfo {eid} {071} (\bibinfo {year} {2022})},\ \Eprint {https://arxiv.org/abs/2201.01756} {arXiv:2201.01756 [hep-ph]} \BibitemShut {NoStop}%
\bibitem [{\citenamefont {{Ofengeim}}\ \emph {et~al.}(2024)\citenamefont {{Ofengeim}}, \citenamefont {{Shternin}},\ and\ \citenamefont {{Piran}}}]{OfPirShtPRD2024}%
  \BibitemOpen
  \bibfield  {author} {\bibinfo {author} {\bibfnamefont {D.~D.}\ \bibnamefont {{Ofengeim}}}, \bibinfo {author} {\bibfnamefont {P.~S.}\ \bibnamefont {{Shternin}}},\ and\ \bibinfo {author} {\bibfnamefont {T.}~\bibnamefont {{Piran}}},\ }\bibfield  {title} {\bibinfo {title} {{Three-parameter characterization of neutron star mass-radius relation and equation of state}},\ }\href {https://doi.org/10.1103/PhysRevD.110.103046} {\bibfield  {journal} {\bibinfo  {journal} {\prd}\ }\textbf {\bibinfo {volume} {110}},\ \bibinfo {eid} {103046} (\bibinfo {year} {2024})},\ \Eprint {https://arxiv.org/abs/2404.17647} {arXiv:2404.17647 [astro-ph.HE]} \BibitemShut {NoStop}%
\bibitem [{\citenamefont {{Sun}}\ and\ \citenamefont {{Lattimer}}(2025)}]{SunLattimerApJ2025}%
  \BibitemOpen
  \bibfield  {author} {\bibinfo {author} {\bibfnamefont {B.}~\bibnamefont {{Sun}}}\ and\ \bibinfo {author} {\bibfnamefont {J.~M.}\ \bibnamefont {{Lattimer}}},\ }\bibfield  {title} {\bibinfo {title} {{Correlations between the Neutron Star Mass{\textendash}Radius Relation and the Equation of State of Dense Matter}},\ }\href {https://doi.org/10.3847/1538-4357/adc25d} {\bibfield  {journal} {\bibinfo  {journal} {\apj}\ }\textbf {\bibinfo {volume} {984}},\ \bibinfo {eid} {30} (\bibinfo {year} {2025})},\ \Eprint {https://arxiv.org/abs/2412.14645} {arXiv:2412.14645 [astro-ph.SR]} \BibitemShut {NoStop}%
\bibitem [{\citenamefont {{Keller}}\ \emph {et~al.}(2023)\citenamefont {{Keller}}, \citenamefont {{Hebeler}},\ and\ \citenamefont {{Schwenk}}}]{Keller+PRL2023}%
  \BibitemOpen
  \bibfield  {author} {\bibinfo {author} {\bibfnamefont {J.}~\bibnamefont {{Keller}}}, \bibinfo {author} {\bibfnamefont {K.}~\bibnamefont {{Hebeler}}},\ and\ \bibinfo {author} {\bibfnamefont {A.}~\bibnamefont {{Schwenk}}},\ }\bibfield  {title} {\bibinfo {title} {{Nuclear Equation of State for Arbitrary Proton Fraction and Temperature Based on Chiral Effective Field Theory and a Gaussian Process Emulator}},\ }\href {https://doi.org/10.1103/PhysRevLett.130.072701} {\bibfield  {journal} {\bibinfo  {journal} {\prl}\ }\textbf {\bibinfo {volume} {130}},\ \bibinfo {eid} {072701} (\bibinfo {year} {2023})},\ \Eprint {https://arxiv.org/abs/2204.14016} {arXiv:2204.14016 [nucl-th]} \BibitemShut {NoStop}%
\bibitem [{\citenamefont {{Komoltsev}}\ and\ \citenamefont {{Kurkela}}(2022)}]{KomoltsevKurkelaPRL2022}%
  \BibitemOpen
  \bibfield  {author} {\bibinfo {author} {\bibfnamefont {O.}~\bibnamefont {{Komoltsev}}}\ and\ \bibinfo {author} {\bibfnamefont {A.}~\bibnamefont {{Kurkela}}},\ }\bibfield  {title} {\bibinfo {title} {{How Perturbative QCD Constrains the Equation of State at Neutron-Star Densities}},\ }\href {https://doi.org/10.1103/PhysRevLett.128.202701} {\bibfield  {journal} {\bibinfo  {journal} {\prl}\ }\textbf {\bibinfo {volume} {128}},\ \bibinfo {eid} {202701} (\bibinfo {year} {2022})},\ \Eprint {https://arxiv.org/abs/2111.05350} {arXiv:2111.05350 [nucl-th]} \BibitemShut {NoStop}%
\bibitem [{\citenamefont {{Ofengeim}}(2020)}]{Ofengeim2020}%
  \BibitemOpen
  \bibfield  {author} {\bibinfo {author} {\bibfnamefont {D.~D.}\ \bibnamefont {{Ofengeim}}},\ }\bibfield  {title} {\bibinfo {title} {{Universal properties of maximum-mass neutron stars: A new tool to explore superdense matter}},\ }\href {https://doi.org/10.1103/PhysRevD.101.103029} {\bibfield  {journal} {\bibinfo  {journal} {\prd}\ }\textbf {\bibinfo {volume} {101}},\ \bibinfo {eid} {103029} (\bibinfo {year} {2020})},\ \Eprint {https://arxiv.org/abs/2005.03549} {arXiv:2005.03549 [astro-ph.HE]} \BibitemShut {NoStop}%
\bibitem [{\citenamefont {{Ofengeim}}\ \emph {et~al.}(2023)\citenamefont {{Ofengeim}}, \citenamefont {{Shternin}},\ and\ \citenamefont {{Piran}}}]{OfShtPirAstLett2023}%
  \BibitemOpen
  \bibfield  {author} {\bibinfo {author} {\bibfnamefont {D.~D.}\ \bibnamefont {{Ofengeim}}}, \bibinfo {author} {\bibfnamefont {P.~S.}\ \bibnamefont {{Shternin}}},\ and\ \bibinfo {author} {\bibfnamefont {T.}~\bibnamefont {{Piran}}},\ }\bibfield  {title} {\bibinfo {title} {{Maximal Mass Neutron Star as a Key to Superdense Matter Physics}},\ }\href {https://doi.org/10.1134/S1063773723100055} {\bibfield  {journal} {\bibinfo  {journal} {Astronomy Letters}\ }\textbf {\bibinfo {volume} {49}},\ \bibinfo {pages} {567} (\bibinfo {year} {2023})},\ \Eprint {https://arxiv.org/abs/2310.16847} {arXiv:2310.16847 [astro-ph.HE]} \BibitemShut {NoStop}%
\bibitem [{\citenamefont {{Cai}}\ \emph {et~al.}(2023)\citenamefont {{Cai}}, \citenamefont {{Li}},\ and\ \citenamefont {{Zhang}}}]{Cai+ApJ2023}%
  \BibitemOpen
  \bibfield  {author} {\bibinfo {author} {\bibfnamefont {B.-J.}\ \bibnamefont {{Cai}}}, \bibinfo {author} {\bibfnamefont {B.-A.}\ \bibnamefont {{Li}}},\ and\ \bibinfo {author} {\bibfnamefont {Z.}~\bibnamefont {{Zhang}}},\ }\bibfield  {title} {\bibinfo {title} {{Core States of Neutron Stars from Anatomizing Their Scaled Structure Equations}},\ }\href {https://doi.org/10.3847/1538-4357/acdef0} {\bibfield  {journal} {\bibinfo  {journal} {\apj}\ }\textbf {\bibinfo {volume} {952}},\ \bibinfo {eid} {147} (\bibinfo {year} {2023})},\ \Eprint {https://arxiv.org/abs/2306.08202} {arXiv:2306.08202 [nucl-th]} \BibitemShut {NoStop}%
\bibitem [{\citenamefont {{Demorest}}\ \emph {et~al.}(2010)\citenamefont {{Demorest}}, \citenamefont {{Pennucci}}, \citenamefont {{Ransom}}, \citenamefont {{Roberts}}, \citenamefont {{Hessels}},\ and\ \citenamefont {et~al.}}]{Demorest+Nat2010_1614}%
  \BibitemOpen
  \bibfield  {author} {\bibinfo {author} {\bibfnamefont {P.~B.}\ \bibnamefont {{Demorest}}}, \bibinfo {author} {\bibfnamefont {T.}~\bibnamefont {{Pennucci}}}, \bibinfo {author} {\bibfnamefont {S.~M.}\ \bibnamefont {{Ransom}}}, \bibinfo {author} {\bibfnamefont {M.~S.~E.}\ \bibnamefont {{Roberts}}}, \bibinfo {author} {\bibfnamefont {J.~W.~T.}\ \bibnamefont {{Hessels}}},\ and\ \bibinfo {author} {\bibnamefont {et~al.}},\ }\bibfield  {title} {\bibinfo {title} {{A two-solar-mass neutron star measured using Shapiro delay}},\ }\href {https://doi.org/10.1038/nature09466} {\bibfield  {journal} {\bibinfo  {journal} {\nat}\ }\textbf {\bibinfo {volume} {467}},\ \bibinfo {pages} {1081} (\bibinfo {year} {2010})},\ \Eprint {https://arxiv.org/abs/1010.5788} {arXiv:1010.5788 [astro-ph.HE]} \BibitemShut {NoStop}%
\bibitem [{\citenamefont {{Agazie}}\ \emph {et~al.}(2023)\citenamefont {{Agazie}}, \citenamefont {{Alam}}, \citenamefont {{Anumarlapudi}}, \citenamefont {{Archibald}}, \citenamefont {{Arzoumanian}},\ and\ \citenamefont {et~al.}}]{NANOGrav15years}%
  \BibitemOpen
  \bibfield  {author} {\bibinfo {author} {\bibfnamefont {G.}~\bibnamefont {{Agazie}}}, \bibinfo {author} {\bibfnamefont {M.~F.}\ \bibnamefont {{Alam}}}, \bibinfo {author} {\bibfnamefont {A.}~\bibnamefont {{Anumarlapudi}}}, \bibinfo {author} {\bibfnamefont {A.~M.}\ \bibnamefont {{Archibald}}}, \bibinfo {author} {\bibfnamefont {Z.}~\bibnamefont {{Arzoumanian}}},\ and\ \bibinfo {author} {\bibnamefont {et~al.}},\ }\bibfield  {title} {\bibinfo {title} {{The NANOGrav 15 yr Data Set: Observations and Timing of 68 Millisecond Pulsars}},\ }\href {https://doi.org/10.3847/2041-8213/acda9a} {\bibfield  {journal} {\bibinfo  {journal} {\apjl}\ }\textbf {\bibinfo {volume} {951}},\ \bibinfo {eid} {L9} (\bibinfo {year} {2023})},\ \Eprint {https://arxiv.org/abs/2306.16217} {arXiv:2306.16217 [astro-ph.HE]} \BibitemShut {NoStop}%
\bibitem [{\citenamefont {{Antoniadis}}\ \emph {et~al.}(2013)\citenamefont {{Antoniadis}}, \citenamefont {{Freire}}, \citenamefont {{Wex}}, \citenamefont {{Tauris}}, \citenamefont {{Lynch}},\ and\ \citenamefont {et~al.}}]{Antoniadis+Sci2013_0348}%
  \BibitemOpen
  \bibfield  {author} {\bibinfo {author} {\bibfnamefont {J.}~\bibnamefont {{Antoniadis}}}, \bibinfo {author} {\bibfnamefont {P.~C.~C.}\ \bibnamefont {{Freire}}}, \bibinfo {author} {\bibfnamefont {N.}~\bibnamefont {{Wex}}}, \bibinfo {author} {\bibfnamefont {T.~M.}\ \bibnamefont {{Tauris}}}, \bibinfo {author} {\bibfnamefont {R.~S.}\ \bibnamefont {{Lynch}}},\ and\ \bibinfo {author} {\bibnamefont {et~al.}},\ }\bibfield  {title} {\bibinfo {title} {{A Massive Pulsar in a Compact Relativistic Binary}},\ }\href {https://doi.org/10.1126/science.1233232} {\bibfield  {journal} {\bibinfo  {journal} {Science}\ }\textbf {\bibinfo {volume} {340}},\ \bibinfo {pages} {448} (\bibinfo {year} {2013})},\ \Eprint {https://arxiv.org/abs/1304.6875} {arXiv:1304.6875 [astro-ph.HE]} \BibitemShut {NoStop}%
\bibitem [{\citenamefont {{Romani}}\ \emph {et~al.}(2022)\citenamefont {{Romani}}, \citenamefont {{Kandel}}, \citenamefont {{Filippenko}}, \citenamefont {{Brink}}, \citenamefont {{Zheng}},\ and\ \citenamefont {et~al.}}]{Romani+ApJL2022_0952}%
  \BibitemOpen
  \bibfield  {author} {\bibinfo {author} {\bibfnamefont {R.~W.}\ \bibnamefont {{Romani}}}, \bibinfo {author} {\bibfnamefont {D.}~\bibnamefont {{Kandel}}}, \bibinfo {author} {\bibfnamefont {A.~V.}\ \bibnamefont {{Filippenko}}}, \bibinfo {author} {\bibfnamefont {T.~G.}\ \bibnamefont {{Brink}}}, \bibinfo {author} {\bibfnamefont {W.}~\bibnamefont {{Zheng}}},\ and\ \bibinfo {author} {\bibnamefont {et~al.}},\ }\bibfield  {title} {\bibinfo {title} {{PSR J0952-0607: The Fastest and Heaviest Known Galactic Neutron Star}},\ }\href {https://doi.org/10.3847/2041-8213/ac8007} {\bibfield  {journal} {\bibinfo  {journal} {\apjl}\ }\textbf {\bibinfo {volume} {934}},\ \bibinfo {eid} {L17} (\bibinfo {year} {2022})},\ \Eprint {https://arxiv.org/abs/2207.05124} {arXiv:2207.05124 [astro-ph.HE]} \BibitemShut {NoStop}%
\bibitem [{\citenamefont {{Romani}}\ \emph {et~al.}(2026)\citenamefont {{Romani}}, \citenamefont {{Beleznay}}, \citenamefont {{Filippenko}}, \citenamefont {{Brink}}, \citenamefont {{Zheng}},\ and\ \citenamefont {et~al.}}]{Romani+2026_0952}%
  \BibitemOpen
  \bibfield  {author} {\bibinfo {author} {\bibfnamefont {R.~W.}\ \bibnamefont {{Romani}}}, \bibinfo {author} {\bibfnamefont {M.}~\bibnamefont {{Beleznay}}}, \bibinfo {author} {\bibfnamefont {A.~V.}\ \bibnamefont {{Filippenko}}}, \bibinfo {author} {\bibfnamefont {T.~G.}\ \bibnamefont {{Brink}}}, \bibinfo {author} {\bibfnamefont {W.}~\bibnamefont {{Zheng}}},\ and\ \bibinfo {author} {\bibnamefont {et~al.}},\ }\bibfield  {title} {\bibinfo {title} {{PSR J0952-0607: Tightening a Record-high Neutron Star Mass}},\ }\href {https://doi.org/10.3847/1538-4357/ae28c5} {\bibfield  {journal} {\bibinfo  {journal} {\apj}\ }\textbf {\bibinfo {volume} {996}},\ \bibinfo {eid} {101} (\bibinfo {year} {2026})},\ \Eprint {https://arxiv.org/abs/2512.05099} {arXiv:2512.05099 [astro-ph.HE]} \BibitemShut {NoStop}%
\bibitem [{\citenamefont {{Rezzolla}}\ \emph {et~al.}(2018)\citenamefont {{Rezzolla}}, \citenamefont {{Most}},\ and\ \citenamefont {{Weih}}}]{RezzollaMostWeihApJL2018}%
  \BibitemOpen
  \bibfield  {author} {\bibinfo {author} {\bibfnamefont {L.}~\bibnamefont {{Rezzolla}}}, \bibinfo {author} {\bibfnamefont {E.~R.}\ \bibnamefont {{Most}}},\ and\ \bibinfo {author} {\bibfnamefont {L.~R.}\ \bibnamefont {{Weih}}},\ }\bibfield  {title} {\bibinfo {title} {{Using Gravitational-wave Observations and Quasi-universal Relations to Constrain the Maximum Mass of Neutron Stars}},\ }\href {https://doi.org/10.3847/2041-8213/aaa401} {\bibfield  {journal} {\bibinfo  {journal} {\apjl}\ }\textbf {\bibinfo {volume} {852}},\ \bibinfo {eid} {L25} (\bibinfo {year} {2018})},\ \Eprint {https://arxiv.org/abs/1711.00314} {arXiv:1711.00314 [astro-ph.HE]} \BibitemShut {NoStop}%
\bibitem [{\citenamefont {{Nathanail}}\ \emph {et~al.}(2021)\citenamefont {{Nathanail}}, \citenamefont {{Most}},\ and\ \citenamefont {{Rezzolla}}}]{NathanailMostRezzollaApJL2021}%
  \BibitemOpen
  \bibfield  {author} {\bibinfo {author} {\bibfnamefont {A.}~\bibnamefont {{Nathanail}}}, \bibinfo {author} {\bibfnamefont {E.~R.}\ \bibnamefont {{Most}}},\ and\ \bibinfo {author} {\bibfnamefont {L.}~\bibnamefont {{Rezzolla}}},\ }\bibfield  {title} {\bibinfo {title} {{GW170817 and GW190814: Tension on the Maximum Mass}},\ }\href {https://doi.org/10.3847/2041-8213/abdfc6} {\bibfield  {journal} {\bibinfo  {journal} {\apjl}\ }\textbf {\bibinfo {volume} {908}},\ \bibinfo {eid} {L28} (\bibinfo {year} {2021})},\ \Eprint {https://arxiv.org/abs/2101.01735} {arXiv:2101.01735 [astro-ph.HE]} \BibitemShut {NoStop}%
\bibitem [{\citenamefont {{Riley}}\ \emph {et~al.}(2021)\citenamefont {{Riley}}, \citenamefont {{Watts}}, \citenamefont {{Ray}}, \citenamefont {{Bogdanov}}, \citenamefont {{Guillot}},\ and\ \citenamefont {et~al.}}]{Riley+ApJL2021}%
  \BibitemOpen
  \bibfield  {author} {\bibinfo {author} {\bibfnamefont {T.~E.}\ \bibnamefont {{Riley}}}, \bibinfo {author} {\bibfnamefont {A.~L.}\ \bibnamefont {{Watts}}}, \bibinfo {author} {\bibfnamefont {P.~S.}\ \bibnamefont {{Ray}}}, \bibinfo {author} {\bibfnamefont {S.}~\bibnamefont {{Bogdanov}}}, \bibinfo {author} {\bibfnamefont {S.}~\bibnamefont {{Guillot}}},\ and\ \bibinfo {author} {\bibnamefont {et~al.}},\ }\bibfield  {title} {\bibinfo {title} {{A NICER View of the Massive Pulsar PSR J0740+6620 Informed by Radio Timing and XMM-Newton Spectroscopy}},\ }\href {https://doi.org/10.3847/2041-8213/ac0a81} {\bibfield  {journal} {\bibinfo  {journal} {\apjl}\ }\textbf {\bibinfo {volume} {918}},\ \bibinfo {eid} {L27} (\bibinfo {year} {2021})},\ \Eprint {https://arxiv.org/abs/2105.06980} {arXiv:2105.06980 [astro-ph.HE]} \BibitemShut {NoStop}%
\bibitem [{\citenamefont {{Salmi}}\ \emph {et~al.}(2024{\natexlab{a}})\citenamefont {{Salmi}}, \citenamefont {{Choudhury}}, \citenamefont {{Kini}}, \citenamefont {{Riley}}, \citenamefont {{Vinciguerra}},\ and\ \citenamefont {et~al.}}]{Salmi+ApJ2024_0740}%
  \BibitemOpen
  \bibfield  {author} {\bibinfo {author} {\bibfnamefont {T.}~\bibnamefont {{Salmi}}}, \bibinfo {author} {\bibfnamefont {D.}~\bibnamefont {{Choudhury}}}, \bibinfo {author} {\bibfnamefont {Y.}~\bibnamefont {{Kini}}}, \bibinfo {author} {\bibfnamefont {T.~E.}\ \bibnamefont {{Riley}}}, \bibinfo {author} {\bibfnamefont {S.}~\bibnamefont {{Vinciguerra}}},\ and\ \bibinfo {author} {\bibnamefont {et~al.}},\ }\bibfield  {title} {\bibinfo {title} {{The Radius of the High-mass Pulsar PSR J0740+6620 with 3.6 yr of NICER Data}},\ }\href {https://doi.org/10.3847/1538-4357/ad5f1f} {\bibfield  {journal} {\bibinfo  {journal} {\apj}\ }\textbf {\bibinfo {volume} {974}},\ \bibinfo {eid} {294} (\bibinfo {year} {2024}{\natexlab{a}})},\ \Eprint {https://arxiv.org/abs/2406.14466} {arXiv:2406.14466 [astro-ph.HE]} \BibitemShut {NoStop}%
\bibitem [{\citenamefont {{Miller}}\ \emph {et~al.}(2019)\citenamefont {{Miller}}, \citenamefont {{Lamb}}, \citenamefont {{Dittmann}}, \citenamefont {{Bogdanov}}, \citenamefont {{Arzoumanian}}, \citenamefont {{Gendreau}}, \citenamefont {{Guillot}}, \citenamefont {{Harding}}, \citenamefont {{Ho}}, \citenamefont {{Lattimer}}, \citenamefont {{Ludlam}}, \citenamefont {{Mahmoodifar}}, \citenamefont {{Morsink}}, \citenamefont {{Ray}}, \citenamefont {{Strohmayer}}, \citenamefont {{Wood}}, \citenamefont {{Enoto}}, \citenamefont {{Foster}}, \citenamefont {{Okajima}}, \citenamefont {{Prigozhin}},\ and\ \citenamefont {{Soong}}}]{Miller+2019ApJ0030}%
  \BibitemOpen
  \bibfield  {author} {\bibinfo {author} {\bibfnamefont {M.~C.}\ \bibnamefont {{Miller}}}, \bibinfo {author} {\bibfnamefont {F.~K.}\ \bibnamefont {{Lamb}}}, \bibinfo {author} {\bibfnamefont {A.~J.}\ \bibnamefont {{Dittmann}}}, \bibinfo {author} {\bibfnamefont {S.}~\bibnamefont {{Bogdanov}}}, \bibinfo {author} {\bibfnamefont {Z.}~\bibnamefont {{Arzoumanian}}}, \bibinfo {author} {\bibfnamefont {K.~C.}\ \bibnamefont {{Gendreau}}}, \bibinfo {author} {\bibfnamefont {S.}~\bibnamefont {{Guillot}}}, \bibinfo {author} {\bibfnamefont {A.~K.}\ \bibnamefont {{Harding}}}, \bibinfo {author} {\bibfnamefont {W.~C.~G.}\ \bibnamefont {{Ho}}}, \bibinfo {author} {\bibfnamefont {J.~M.}\ \bibnamefont {{Lattimer}}}, \bibinfo {author} {\bibfnamefont {R.~M.}\ \bibnamefont {{Ludlam}}}, \bibinfo {author} {\bibfnamefont {S.}~\bibnamefont {{Mahmoodifar}}}, \bibinfo {author} {\bibfnamefont {S.~M.}\ \bibnamefont {{Morsink}}}, \bibinfo {author} {\bibfnamefont {P.~S.}\ \bibnamefont {{Ray}}}, \bibinfo {author} {\bibfnamefont {T.~E.}\
  \bibnamefont {{Strohmayer}}}, \bibinfo {author} {\bibfnamefont {K.~S.}\ \bibnamefont {{Wood}}}, \bibinfo {author} {\bibfnamefont {T.}~\bibnamefont {{Enoto}}}, \bibinfo {author} {\bibfnamefont {R.}~\bibnamefont {{Foster}}}, \bibinfo {author} {\bibfnamefont {T.}~\bibnamefont {{Okajima}}}, \bibinfo {author} {\bibfnamefont {G.}~\bibnamefont {{Prigozhin}}},\ and\ \bibinfo {author} {\bibfnamefont {Y.}~\bibnamefont {{Soong}}},\ }\bibfield  {title} {\bibinfo {title} {{PSR J0030+0451 Mass and Radius from NICER Data and Implications for the Properties of Neutron Star Matter}},\ }\href {https://doi.org/10.3847/2041-8213/ab50c5} {\bibfield  {journal} {\bibinfo  {journal} {\apjl}\ }\textbf {\bibinfo {volume} {887}},\ \bibinfo {eid} {L24} (\bibinfo {year} {2019})},\ \Eprint {https://arxiv.org/abs/1912.05705} {arXiv:1912.05705 [astro-ph.HE]} \BibitemShut {NoStop}%
\bibitem [{\citenamefont {{Vinciguerra}}\ \emph {et~al.}(2024)\citenamefont {{Vinciguerra}}, \citenamefont {{Salmi}}, \citenamefont {{Watts}}, \citenamefont {{Choudhury}}, \citenamefont {{Riley}},\ and\ \citenamefont {et~al.}}]{Vinciguerra+ApJ2024}%
  \BibitemOpen
  \bibfield  {author} {\bibinfo {author} {\bibfnamefont {S.}~\bibnamefont {{Vinciguerra}}}, \bibinfo {author} {\bibfnamefont {T.}~\bibnamefont {{Salmi}}}, \bibinfo {author} {\bibfnamefont {A.~L.}\ \bibnamefont {{Watts}}}, \bibinfo {author} {\bibfnamefont {D.}~\bibnamefont {{Choudhury}}}, \bibinfo {author} {\bibfnamefont {T.~E.}\ \bibnamefont {{Riley}}},\ and\ \bibinfo {author} {\bibnamefont {et~al.}},\ }\bibfield  {title} {\bibinfo {title} {{An Updated Mass-Radius Analysis of the 2017-2018 NICER Data Set of PSR J0030+0451}},\ }\href {https://doi.org/10.3847/1538-4357/acfb83} {\bibfield  {journal} {\bibinfo  {journal} {\apj}\ }\textbf {\bibinfo {volume} {961}},\ \bibinfo {eid} {62} (\bibinfo {year} {2024})},\ \Eprint {https://arxiv.org/abs/2308.09469} {arXiv:2308.09469 [astro-ph.HE]} \BibitemShut {NoStop}%
\bibitem [{\citenamefont {{Saffer}}\ \emph {et~al.}(2025)\citenamefont {{Saffer}}, \citenamefont {{Fonseca}}, \citenamefont {{Ransom}}, \citenamefont {{Stairs}}, \citenamefont {{Lynch}},\ and\ \citenamefont {et~al.}}]{Saffer+ApJL2025_0348}%
  \BibitemOpen
  \bibfield  {author} {\bibinfo {author} {\bibfnamefont {A.}~\bibnamefont {{Saffer}}}, \bibinfo {author} {\bibfnamefont {E.}~\bibnamefont {{Fonseca}}}, \bibinfo {author} {\bibfnamefont {S.}~\bibnamefont {{Ransom}}}, \bibinfo {author} {\bibfnamefont {I.}~\bibnamefont {{Stairs}}}, \bibinfo {author} {\bibfnamefont {R.}~\bibnamefont {{Lynch}}},\ and\ \bibinfo {author} {\bibnamefont {et~al.}},\ }\bibfield  {title} {\bibinfo {title} {{A Lower Mass Estimate for PSR J0348+0432 Based on CHIME/Pulsar Precision Timing}},\ }\href {https://doi.org/10.3847/2041-8213/adc25e} {\bibfield  {journal} {\bibinfo  {journal} {\apjl}\ }\textbf {\bibinfo {volume} {983}},\ \bibinfo {eid} {L20} (\bibinfo {year} {2025})},\ \Eprint {https://arxiv.org/abs/2412.02850} {arXiv:2412.02850 [astro-ph.HE]} \BibitemShut {NoStop}%
\bibitem [{\citenamefont {{Sullivan}}\ and\ \citenamefont {{Romani}}(2024)}]{SullivanRomaniApJ2024_2215}%
  \BibitemOpen
  \bibfield  {author} {\bibinfo {author} {\bibfnamefont {A.~G.}\ \bibnamefont {{Sullivan}}}\ and\ \bibinfo {author} {\bibfnamefont {R.~W.}\ \bibnamefont {{Romani}}},\ }\bibfield  {title} {\bibinfo {title} {{The Intrabinary Shock and Companion Star of Redback Pulsar J2215+5135}},\ }\href {https://doi.org/10.3847/1538-4357/ad4d85} {\bibfield  {journal} {\bibinfo  {journal} {\apj}\ }\textbf {\bibinfo {volume} {974}},\ \bibinfo {eid} {315} (\bibinfo {year} {2024})},\ \Eprint {https://arxiv.org/abs/2405.13889} {arXiv:2405.13889 [astro-ph.HE]} \BibitemShut {NoStop}%
\bibitem [{\citenamefont {{Mauviard}}\ \emph {et~al.}(2025)\citenamefont {{Mauviard}}, \citenamefont {{Guillot}}, \citenamefont {{Salmi}}, \citenamefont {{Choudhury}}, \citenamefont {{Dorsman}},\ and\ \citenamefont {et~al.}}]{Mauviard+2025_0614}%
  \BibitemOpen
  \bibfield  {author} {\bibinfo {author} {\bibfnamefont {L.}~\bibnamefont {{Mauviard}}}, \bibinfo {author} {\bibfnamefont {S.}~\bibnamefont {{Guillot}}}, \bibinfo {author} {\bibfnamefont {T.}~\bibnamefont {{Salmi}}}, \bibinfo {author} {\bibfnamefont {D.}~\bibnamefont {{Choudhury}}}, \bibinfo {author} {\bibfnamefont {B.}~\bibnamefont {{Dorsman}}},\ and\ \bibinfo {author} {\bibnamefont {et~al.}},\ }\bibfield  {title} {\bibinfo {title} {{A NICER View of the 1.4 M$_{{\ensuremath{\odot}}}$ Edge-on Pulsar PSR J0614-3329}},\ }\href {https://doi.org/10.3847/1538-4357/ae145d} {\bibfield  {journal} {\bibinfo  {journal} {\apj}\ }\textbf {\bibinfo {volume} {995}},\ \bibinfo {eid} {60} (\bibinfo {year} {2025})},\ \Eprint {https://arxiv.org/abs/2506.14883} {arXiv:2506.14883 [astro-ph.HE]} \BibitemShut {NoStop}%
\bibitem [{\citenamefont {{Salmi}}\ \emph {et~al.}(2024{\natexlab{b}})\citenamefont {{Salmi}}, \citenamefont {{Deneva}}, \citenamefont {{Ray}}, \citenamefont {{Watts}}, \citenamefont {{Choudhury}},\ and\ \citenamefont {et~al.}}]{Salmi+ApJ2024_1231}%
  \BibitemOpen
  \bibfield  {author} {\bibinfo {author} {\bibfnamefont {T.}~\bibnamefont {{Salmi}}}, \bibinfo {author} {\bibfnamefont {J.~S.}\ \bibnamefont {{Deneva}}}, \bibinfo {author} {\bibfnamefont {P.~S.}\ \bibnamefont {{Ray}}}, \bibinfo {author} {\bibfnamefont {A.~L.}\ \bibnamefont {{Watts}}}, \bibinfo {author} {\bibfnamefont {D.}~\bibnamefont {{Choudhury}}},\ and\ \bibinfo {author} {\bibnamefont {et~al.}},\ }\bibfield  {title} {\bibinfo {title} {{A NICER View of PSR J1231{\ensuremath{-}}1411: A Complex Case}},\ }\href {https://doi.org/10.3847/1538-4357/ad81d2} {\bibfield  {journal} {\bibinfo  {journal} {\apj}\ }\textbf {\bibinfo {volume} {976}},\ \bibinfo {eid} {58} (\bibinfo {year} {2024}{\natexlab{b}})},\ \Eprint {https://arxiv.org/abs/2409.14923} {arXiv:2409.14923 [astro-ph.HE]} \BibitemShut {NoStop}%
\bibitem [{\citenamefont {{Abbott}}\ \emph {et~al.}(2017)\citenamefont {{Abbott}}, \citenamefont {{Abbott}}, \citenamefont {{Abbott}} \emph {et~al.}}]{GW170817}%
  \BibitemOpen
  \bibfield  {author} {\bibinfo {author} {\bibfnamefont {B.~P.}\ \bibnamefont {{Abbott}}}, \bibinfo {author} {\bibfnamefont {R.}~\bibnamefont {{Abbott}}}, \bibinfo {author} {\bibfnamefont {T.~D.}\ \bibnamefont {{Abbott}}}, \emph {et~al.},\ }\bibfield  {title} {\bibinfo {title} {{GW170817: Observation of Gravitational Waves from a Binary Neutron Star Inspiral}},\ }\href {https://doi.org/10.1103/PhysRevLett.119.161101} {\bibfield  {journal} {\bibinfo  {journal} {\prl}\ }\textbf {\bibinfo {volume} {119}},\ \bibinfo {eid} {161101} (\bibinfo {year} {2017})},\ \Eprint {https://arxiv.org/abs/1710.05832} {arXiv:1710.05832 [gr-qc]} \BibitemShut {NoStop}%
\bibitem [{\citenamefont {{Abbott}}\ \emph {et~al.}(2020{\natexlab{a}})\citenamefont {{Abbott}}, \citenamefont {{Abbott}}, \citenamefont {{Abbott}}, \citenamefont {{Abraham}}, \citenamefont {{Acernese}}, \citenamefont {{Ackley}}, \citenamefont {{Adams}}, \citenamefont {{Adhikari}}, \citenamefont {{Adya}}, \citenamefont {{Affeldt}},\ and\ \citenamefont {et~al.}}]{GW190425_2020ApJ}%
  \BibitemOpen
  \bibfield  {author} {\bibinfo {author} {\bibfnamefont {B.~P.}\ \bibnamefont {{Abbott}}}, \bibinfo {author} {\bibfnamefont {R.}~\bibnamefont {{Abbott}}}, \bibinfo {author} {\bibfnamefont {T.~D.}\ \bibnamefont {{Abbott}}}, \bibinfo {author} {\bibfnamefont {S.}~\bibnamefont {{Abraham}}}, \bibinfo {author} {\bibfnamefont {F.}~\bibnamefont {{Acernese}}}, \bibinfo {author} {\bibfnamefont {K.}~\bibnamefont {{Ackley}}}, \bibinfo {author} {\bibfnamefont {C.}~\bibnamefont {{Adams}}}, \bibinfo {author} {\bibfnamefont {R.~X.}\ \bibnamefont {{Adhikari}}}, \bibinfo {author} {\bibfnamefont {V.~B.}\ \bibnamefont {{Adya}}}, \bibinfo {author} {\bibfnamefont {C.}~\bibnamefont {{Affeldt}}},\ and\ \bibinfo {author} {\bibnamefont {et~al.}},\ }\bibfield  {title} {\bibinfo {title} {{GW190425: Observation of a Compact Binary Coalescence with Total Mass {\ensuremath{\sim}} 3.4 M$_{{\ensuremath{\odot}}}$}},\ }\href {https://doi.org/10.3847/2041-8213/ab75f5} {\bibfield  {journal} {\bibinfo  {journal} {\apjl}\ }\textbf {\bibinfo
  {volume} {892}},\ \bibinfo {eid} {L3} (\bibinfo {year} {2020}{\natexlab{a}})},\ \Eprint {https://arxiv.org/abs/2001.01761} {arXiv:2001.01761 [astro-ph.HE]} \BibitemShut {NoStop}%
\bibitem [{\citenamefont {{Abbott}}\ \emph {et~al.}(2020{\natexlab{b}})\citenamefont {{Abbott}}, \citenamefont {{Abbott}}, \citenamefont {{Abraham}}, \citenamefont {{Acernese}}, \citenamefont {{Ackley}}, \citenamefont {{Adams}}, \citenamefont {{Adhikari}}, \citenamefont {{Adya}}, \citenamefont {{Affeldt}}, \citenamefont {{Agathos}},\ and\ \citenamefont {et~al.}}]{GW190814_2020ApJ}%
  \BibitemOpen
  \bibfield  {author} {\bibinfo {author} {\bibfnamefont {R.}~\bibnamefont {{Abbott}}}, \bibinfo {author} {\bibfnamefont {T.~D.}\ \bibnamefont {{Abbott}}}, \bibinfo {author} {\bibfnamefont {S.}~\bibnamefont {{Abraham}}}, \bibinfo {author} {\bibfnamefont {F.}~\bibnamefont {{Acernese}}}, \bibinfo {author} {\bibfnamefont {K.}~\bibnamefont {{Ackley}}}, \bibinfo {author} {\bibfnamefont {C.}~\bibnamefont {{Adams}}}, \bibinfo {author} {\bibfnamefont {R.~X.}\ \bibnamefont {{Adhikari}}}, \bibinfo {author} {\bibfnamefont {V.~B.}\ \bibnamefont {{Adya}}}, \bibinfo {author} {\bibfnamefont {C.}~\bibnamefont {{Affeldt}}}, \bibinfo {author} {\bibfnamefont {M.}~\bibnamefont {{Agathos}}},\ and\ \bibinfo {author} {\bibnamefont {et~al.}},\ }\bibfield  {title} {\bibinfo {title} {{GW190814: Gravitational Waves from the Coalescence of a 23 Solar Mass Black Hole with a 2.6 Solar Mass Compact Object}},\ }\href {https://doi.org/10.3847/2041-8213/ab960f} {\bibfield  {journal} {\bibinfo  {journal} {\apjl}\ }\textbf {\bibinfo {volume}
  {896}},\ \bibinfo {eid} {L44} (\bibinfo {year} {2020}{\natexlab{b}})},\ \Eprint {https://arxiv.org/abs/2006.12611} {arXiv:2006.12611 [astro-ph.HE]} \BibitemShut {NoStop}%
\bibitem [{\citenamefont {{Drischler}}\ \emph {et~al.}(2021)\citenamefont {{Drischler}}, \citenamefont {{Han}}, \citenamefont {{Lattimer}}, \citenamefont {{Prakash}}, \citenamefont {{Reddy}},\ and\ \citenamefont {et~al.}}]{Drischler+PRC2021}%
  \BibitemOpen
  \bibfield  {author} {\bibinfo {author} {\bibfnamefont {C.}~\bibnamefont {{Drischler}}}, \bibinfo {author} {\bibfnamefont {S.}~\bibnamefont {{Han}}}, \bibinfo {author} {\bibfnamefont {J.~M.}\ \bibnamefont {{Lattimer}}}, \bibinfo {author} {\bibfnamefont {M.}~\bibnamefont {{Prakash}}}, \bibinfo {author} {\bibfnamefont {S.}~\bibnamefont {{Reddy}}},\ and\ \bibinfo {author} {\bibnamefont {et~al.}},\ }\bibfield  {title} {\bibinfo {title} {{Limiting masses and radii of neutron stars and their implications}},\ }\href {https://doi.org/10.1103/PhysRevC.103.045808} {\bibfield  {journal} {\bibinfo  {journal} {\prc}\ }\textbf {\bibinfo {volume} {103}},\ \bibinfo {eid} {045808} (\bibinfo {year} {2021})},\ \Eprint {https://arxiv.org/abs/2009.06441} {arXiv:2009.06441 [nucl-th]} \BibitemShut {NoStop}%
\bibitem [{\citenamefont {Glendenning}(2000)}]{Glendenning}%
  \BibitemOpen
  \bibinfo {editor} {\bibfnamefont {N.~K.}\ \bibnamefont {Glendenning}},\ ed.,\ \href@noop {} {\emph {\bibinfo {title} {Compact Stars: Nuclear Physics, Particle Physics, and General Relativity}}},\ \bibinfo {edition} {2nd}\ ed.,\ Astronomy and Astrophysics Library\ (\bibinfo  {publisher} {Springer-Verlag},\ \bibinfo {address} {New York},\ \bibinfo {year} {2000})\BibitemShut {NoStop}%
\bibitem [{\citenamefont {{Ayriyan}}\ \emph {et~al.}(2025)\citenamefont {{Ayriyan}}, \citenamefont {{Ivanytskyi}},\ and\ \citenamefont {{Blaschke}}}]{Ayriyan+2025}%
  \BibitemOpen
  \bibfield  {author} {\bibinfo {author} {\bibfnamefont {A.}~\bibnamefont {{Ayriyan}}}, \bibinfo {author} {\bibfnamefont {O.}~\bibnamefont {{Ivanytskyi}}},\ and\ \bibinfo {author} {\bibfnamefont {D.}~\bibnamefont {{Blaschke}}},\ }\bibfield  {title} {\bibinfo {title} {{Bayesian inference favors quark matter in neutron star interiors}},\ }\href {https://doi.org/10.48550/arXiv.2509.02554} {\bibfield  {journal} {\bibinfo  {journal} {arXiv e-prints}\ ,\ \bibinfo {eid} {arXiv:2509.02554}} (\bibinfo {year} {2025})},\ \Eprint {https://arxiv.org/abs/2509.02554} {arXiv:2509.02554 [nucl-th]} \BibitemShut {NoStop}%
\bibitem [{\citenamefont {{Misra}}\ \emph {et~al.}(2025)\citenamefont {{Misra}}, \citenamefont {{Linares}},\ and\ \citenamefont {{Ye}}}]{MisraLinaresYeAA2025_Spiders}%
  \BibitemOpen
  \bibfield  {author} {\bibinfo {author} {\bibfnamefont {D.}~\bibnamefont {{Misra}}}, \bibinfo {author} {\bibfnamefont {M.}~\bibnamefont {{Linares}}},\ and\ \bibinfo {author} {\bibfnamefont {C.~S.}\ \bibnamefont {{Ye}}},\ }\bibfield  {title} {\bibinfo {title} {{Investigating cannibalistic millisecond pulsar binaries using MESA: New constraints from pulsar spin and mass evolution}},\ }\href {https://doi.org/10.1051/0004-6361/202452035} {\bibfield  {journal} {\bibinfo  {journal} {\aap}\ }\textbf {\bibinfo {volume} {693}},\ \bibinfo {eid} {A314} (\bibinfo {year} {2025})},\ \Eprint {https://arxiv.org/abs/2408.16048} {arXiv:2408.16048 [astro-ph.HE]} \BibitemShut {NoStop}%
\bibitem [{\citenamefont {{Perna}}\ \emph {et~al.}(2025)\citenamefont {{Perna}}, \citenamefont {{Gottlieb}}, \citenamefont {{Shukla}},\ and\ \citenamefont {{Radice}}}]{PernaGottlieb+PRD2025}%
  \BibitemOpen
  \bibfield  {author} {\bibinfo {author} {\bibfnamefont {R.}~\bibnamefont {{Perna}}}, \bibinfo {author} {\bibfnamefont {O.}~\bibnamefont {{Gottlieb}}}, \bibinfo {author} {\bibfnamefont {E.}~\bibnamefont {{Shukla}}},\ and\ \bibinfo {author} {\bibfnamefont {D.}~\bibnamefont {{Radice}}},\ }\bibfield  {title} {\bibinfo {title} {{Connecting GRBs from binary neutron star mergers to nuclear properties of neutron stars}},\ }\href {https://doi.org/10.1103/PhysRevD.111.063015} {\bibfield  {journal} {\bibinfo  {journal} {\prd}\ }\textbf {\bibinfo {volume} {111}},\ \bibinfo {eid} {063015} (\bibinfo {year} {2025})},\ \Eprint {https://arxiv.org/abs/2412.07846} {arXiv:2412.07846 [astro-ph.HE]} \BibitemShut {NoStop}%
\bibitem [{\citenamefont {{Chen}}\ and\ \citenamefont {{Gottlieb}}(2025)}]{ChenGottlieb2025_Mmax}%
  \BibitemOpen
  \bibfield  {author} {\bibinfo {author} {\bibfnamefont {H.-Y.}\ \bibnamefont {{Chen}}}\ and\ \bibinfo {author} {\bibfnamefont {O.}~\bibnamefont {{Gottlieb}}},\ }\bibfield  {title} {\bibinfo {title} {{Inferring Neutron Star Nuclear Properties from Gravitational-Wave and Gamma-Ray Burst Observations}},\ }\href {https://doi.org/10.48550/arXiv.2506.18151} {\bibfield  {journal} {\bibinfo  {journal} {arXiv e-prints}\ ,\ \bibinfo {eid} {arXiv:2506.18151}} (\bibinfo {year} {2025})},\ \Eprint {https://arxiv.org/abs/2506.18151} {arXiv:2506.18151 [astro-ph.HE]} \BibitemShut {NoStop}%
\bibitem [{\citenamefont {{Lattimer}}\ and\ \citenamefont {{Schutz}}(2005)}]{LattimerSchutzApJ2005}%
  \BibitemOpen
  \bibfield  {author} {\bibinfo {author} {\bibfnamefont {J.~M.}\ \bibnamefont {{Lattimer}}}\ and\ \bibinfo {author} {\bibfnamefont {B.~F.}\ \bibnamefont {{Schutz}}},\ }\bibfield  {title} {\bibinfo {title} {{Constraining the Equation of State with Moment of Inertia Measurements}},\ }\href {https://doi.org/10.1086/431543} {\bibfield  {journal} {\bibinfo  {journal} {\apj}\ }\textbf {\bibinfo {volume} {629}},\ \bibinfo {pages} {979} (\bibinfo {year} {2005})},\ \Eprint {https://arxiv.org/abs/astro-ph/0411470} {arXiv:astro-ph/0411470 [astro-ph]} \BibitemShut {NoStop}%
\bibitem [{\citenamefont {{Iacovelli}}\ \emph {et~al.}(2023)\citenamefont {{Iacovelli}}, \citenamefont {{Mancarella}}, \citenamefont {{Mondal}}, \citenamefont {{Puecher}}, \citenamefont {{Dietrich}},\ and\ \citenamefont {et~al.}}]{Iacovelli+PRD2023_NucPhysET}%
  \BibitemOpen
  \bibfield  {author} {\bibinfo {author} {\bibfnamefont {F.}~\bibnamefont {{Iacovelli}}}, \bibinfo {author} {\bibfnamefont {M.}~\bibnamefont {{Mancarella}}}, \bibinfo {author} {\bibfnamefont {C.}~\bibnamefont {{Mondal}}}, \bibinfo {author} {\bibfnamefont {A.}~\bibnamefont {{Puecher}}}, \bibinfo {author} {\bibfnamefont {T.}~\bibnamefont {{Dietrich}}},\ and\ \bibinfo {author} {\bibnamefont {et~al.}},\ }\bibfield  {title} {\bibinfo {title} {{Nuclear physics constraints from binary neutron star mergers in the Einstein Telescope era}},\ }\href {https://doi.org/10.1103/PhysRevD.108.122006} {\bibfield  {journal} {\bibinfo  {journal} {\prd}\ }\textbf {\bibinfo {volume} {108}},\ \bibinfo {eid} {122006} (\bibinfo {year} {2023})},\ \Eprint {https://arxiv.org/abs/2308.12378} {arXiv:2308.12378 [gr-qc]} \BibitemShut {NoStop}%
\bibitem [{\citenamefont {{Yagi}}\ and\ \citenamefont {{Yunes}}(2013{\natexlab{b}})}]{YagiYounesSci2013}%
  \BibitemOpen
  \bibfield  {author} {\bibinfo {author} {\bibfnamefont {K.}~\bibnamefont {{Yagi}}}\ and\ \bibinfo {author} {\bibfnamefont {N.}~\bibnamefont {{Yunes}}},\ }\bibfield  {title} {\bibinfo {title} {{I-Love-Q: Unexpected Universal Relations for Neutron Stars and Quark Stars}},\ }\href {https://doi.org/10.1126/science.1236462} {\bibfield  {journal} {\bibinfo  {journal} {Science}\ }\textbf {\bibinfo {volume} {341}},\ \bibinfo {pages} {365} (\bibinfo {year} {2013}{\natexlab{b}})},\ \Eprint {https://arxiv.org/abs/1302.4499} {arXiv:1302.4499 [gr-qc]} \BibitemShut {NoStop}%
\bibitem [{\citenamefont {{Andersson}}\ and\ \citenamefont {{Kokkotas}}(1998)}]{AnderssonKokkotasMNRAS1998}%
  \BibitemOpen
  \bibfield  {author} {\bibinfo {author} {\bibfnamefont {N.}~\bibnamefont {{Andersson}}}\ and\ \bibinfo {author} {\bibfnamefont {K.~D.}\ \bibnamefont {{Kokkotas}}},\ }\bibfield  {title} {\bibinfo {title} {{Towards gravitational wave asteroseismology}},\ }\href {https://doi.org/10.1046/j.1365-8711.1998.01840.x} {\bibfield  {journal} {\bibinfo  {journal} {\mnras}\ }\textbf {\bibinfo {volume} {299}},\ \bibinfo {pages} {1059} (\bibinfo {year} {1998})},\ \Eprint {https://arxiv.org/abs/gr-qc/9711088} {arXiv:gr-qc/9711088 [gr-qc]} \BibitemShut {NoStop}%
\bibitem [{\citenamefont {{Rhoades}}\ and\ \citenamefont {{Ruffini}}(1974)}]{RhoadesRuffini1974}%
  \BibitemOpen
  \bibfield  {author} {\bibinfo {author} {\bibfnamefont {C.~E.}\ \bibnamefont {{Rhoades}}}\ and\ \bibinfo {author} {\bibfnamefont {R.}~\bibnamefont {{Ruffini}}},\ }\bibfield  {title} {\bibinfo {title} {{Maximum Mass of a Neutron Star}},\ }\href {https://doi.org/10.1103/PhysRevLett.32.324} {\bibfield  {journal} {\bibinfo  {journal} {\prl}\ }\textbf {\bibinfo {volume} {32}},\ \bibinfo {pages} {324} (\bibinfo {year} {1974})}\BibitemShut {NoStop}%
\end{thebibliography}
\end{document}